\documentclass{aa} 
\usepackage{graphicx}
\usepackage{lscape}
\usepackage{subfigure}
\usepackage{natbib}
\usepackage{hyperref}
\usepackage[varg]{txfonts}
\usepackage{longtable}
\usepackage{booktabs}
\usepackage[normalem]{ulem} 

\usepackage{xcolor}

\bibpunct{(}{)}{;}{a}{}{,} 
\begin{document}

   \title{Optically variable active galactic nuclei in the 3 yr VST survey of the COSMOS field\thanks{Observations were provided by the ESO programs 088.D-4013, 092.D-0370, and 094.D-0417 (PI G. Pignata).}}

   \author{D. De Cicco \inst{1,2,3}, M. Paolillo\inst{3,4,5}, S. Falocco\inst{6}, M. Poulain\inst{7}, W. N. Brandt\inst{8,9}, F. E. Bauer\inst{2,1,10}, F. Vagnetti\inst{11}, G. Longo\inst{3}, A. Grado\inst{5}, F. Ragosta\inst{3,5}, M. T. Botticella\inst{5}, G. Pignata\inst{12,1}, M. Vaccari\inst{13,14}, M. Radovich\inst{15}, M. Salvato\inst{16}, G. Covone\inst{3,4}, N. R. Napolitano\inst{17}, L. Marchetti\inst{18,13,14}, P. Schipani\inst{5}}
   \authorrunning{D. De Cicco et al.}

\institute{Millennium Institute of Astrophysics (MAS), Nuncio Monse\~nor Sotero Sanz 100, Providencia, Santiago, Chile    
\\e-mail: demetradecicco@gmail.com, ddecicco@astro.puc.cl
  \and
  Instituto de Astrof\'{i}sica, Pontificia Universidad Cat\'{o}lica de Chile, Av. Vicu\~{n}a Mackenna 4860, 7820436 Macul, Santiago, Chile 
  \and
  Department of Physics, University of Napoli ``Federico II'', via Cinthia 9, 80126 Napoli, Italy
    \and
    INFN - Sezione di Napoli, via Cinthia 9, 80126 Napoli, Italy 
 	\and
		INAF - Osservatorio Astronomico di Capodimonte, via Moiariello 16, 80131 Napoli, Italy 
  \and
		KTH Royal Institute of Technology, Brinellv\"{a}gen 8, 114 28 Stockholm, Sweden 
  \and 
        Institut f\"{u}r Astro- und Teilchenphysik, Universit\"{a}t Innsbruck, Technikerstra\ss e 25/8, Innsbruck, A-6020, Austria 
	\and
		Department of Astronomy and Astrophysics, The Pennsylvania State University, University Park, PA 16802, USA 
	\and
		Institute for Gravitation and the Cosmos, The Pennsylvania State University, University Park, PA 16802, USA	
	\and
	    Space Science Institute, 4750 Walnut Street, Suite 2015, Boulder, CO 80301, USA 
	 \and
		Department of Physics, University of Roma ``Tor Vergata'', via della Ricerca Scientifica 1, 00133 Roma, Italy 
	\and
		Departamento de Ciencias Fisicas, Universidad Andres Bello, Avda. Republica 252, Santiago, Chile 
	\and
	 	Department of Physics and Astronomy, University of the Western Cape, Private Bag X17, 7535 Bellville, Cape Town, South Africa 
	\and
		INAF - Istituto di Radioastronomia, via Gobetti 101, 40129 Bologna, Italy 
	\and
		INAF - Osservatorio Astronomico di Padova, vicolo dell'Osservatorio 5, I-35122 Padova, Italy 
	\and
		Max Planck Institut f\"{u}r Extraterrestrische Physik, Giessenbachstra\ss e 1, D-85748 Garching bei M\"{u}nchen, Germany 
	\and 
	    School of Physics and Astronomy, Sun Yat-sen University, Guangzhou 519082, Zhuhai Campus, P.R. China 
    \and
        Department of Astronomy, University of Cape Town, Private Bag X3, Rondebosch 7701, Cape Town, South Africa 
    \\}

   \date{}
  \abstract
   {The analysis of the variability of active galactic nuclei (AGNs) at different wavelengths and the study of possible correlations among different spectral windows are nowadays a major field of inquiry. Optical variability has been largely used to identify AGNs in multivisit surveys. The strength of a selection based on optical variability lies in the chance to analyze data from surveys of large sky areas by ground-based telescopes. However the effectiveness of optical variability selection, with respect to other multiwavelength techniques, has been poorly studied down to the depth expected from next generation surveys.}
   {Here we present the results of our \emph{r}-band analysis of a sample of 299 optically variable AGN candidates in the VST survey of the COSMOS field, counting 54 visits spread over three observing seasons spanning $>3$ yr. This dataset is > 3 times larger in size than the one presented in our previous analysis \citep{decicco}, and the observing baseline is $\sim 8$ times longer.}
   {We push towards deeper magnitudes (\emph{r}(AB) $\sim 23.5$ mag) compared to past studies; we make wide use of ancillary multiwavelength catalogs in order to confirm the nature of our AGN candidates, and constrain the accuracy of the method based on spectroscopic and photometric diagnostics. We also perform tests aimed at assessing the relevance of dense sampling in view of future wide-field surveys.}
   {We demonstrate that the method allows the selection of high-purity (>86\%) samples. We take advantage of the longer observing baseline to achieve great improvement in the completeness of our sample with respect to X-ray and spectroscopically confirmed samples of AGNs (59\%, vs. $\sim 15\%$ in our previous work), as well as in the completeness of unobscured and obscured AGNs.\\
   The effectiveness of the method confirms the importance to develop future, more refined techniques for the automated analysis of larger datasets.}
   {}


   \maketitle
   
%

\section{Introduction}
\label{section:introduction}
Super-massive black holes (SMBHs) are considered ubiquitous guests in the centers of massive galaxies \citep[e.g.,][]{magorrian,Ferrarese&Merritt}. At some point in their lives, they are powered by accreting matter infall; galaxies experiencing such a phase are known as active galactic nuclei (AGNs).

AGN emission is broadband, covering most of the electromagnetic spectrum, as a result of several continuum emission processes plus emission-line features. These features are typically broader and much more prominent than the ones commonly observed in the spectra of inactive galaxies. 

Variability is a signature of AGN emission, and is observed at all frequencies, spanning more than twenty orders of magnitude \citep[e.g.,][and references therein]{padovani}. It affects both continuum and line emission to an extent depending on a number of factors: observed spectral range, timescale, presence of radio jets, source luminosity, mass, accretion rate, etc. \citep[e.g.,][]{paolillo17}. Variations are irregular and aperiodic; their typical timescales range from hours to years, though variations can occasionally be detected on timescales down to minutes or up to $10^4$ yr (see, e.g., voorwerpjes, \citealt{lintott}). Several works over the past decades have widely investigated AGN variability in different wavebands, together with the correlations between variations in different spectral windows \citep[e.g.,][]{Ulrich, Gaskell&Klimek}. An accurate characterization is nowadays possible thanks to surveys with repeated observations over years/decades. In particular, multivisit surveys have been extensively used to search for unobscured AGNs \citep[e.g.,][and references therein]{mushotzky, Klesman&Sarajedini, Trevese, Schmidt, Villforth1, macleod, sarajedini11, choi, graham}. These surveys are generally characterized by irregular sampling, with observing gaps due to observational constraints. The features of the observed variability depend on several parameters, e.g., baseline, number of visits, observing cadence, chosen band, depth, photometric accuracy. 

The next years will see the advent of new-generation telescopes, designed to survey wide sky areas with a high cadence. Among them, the most highly anticipated is undoubtedly the Large Synoptic Survey Telescope (LSST; see, e.g., \citealt{lsst}). LSST surveys are designed to cover about half the sky. At completion, they will be 10 to 100 times deeper than any present wide-field survey, with hundreds to thousands of time samples. About 90\% of LSST’s time will be dedicated to the main survey; the remaining 10\% will go into several mini-surveys. These will include ultra-deep observations of 5-10 so-called Deep Drilling Fields (DDFs). 
The DDFs will target previously surveyed areas, such as COSMOS and the \emph{Chandra} Deep Field-South (CDFS). For them we can expect up to $\sim 14000$ visits, allowing to reach co-added depths of $ugri \sim 28.5$, $z \sim 28$, and $y \sim 27.5$ mag, over the 10 year survey program. This, combined with the multiwavelength coverage from current and future telescopes, makes the DDFs ideal for AGN studies \citep[see, e.g.,][]{ddf}.
The first formal survey data from the LSST main survey are expected no sooner than 2023 (best-case scenario). In anticipation, the astronomical community can and must prepare to deal with them, developing fast and reliable classification tools that will be needed to adequately face unprecedented data streams. 
In this context, we are putting effort into the development and refinement of an effective technique for AGN selection based on their optical variability, with the aim of applying it to wider datasets from surveys to come. We elaborated and tested our selection method analyzing data from the SUpernova Diversity And Rate Evolution (SUDARE; \citealt{botticella13}) survey by the VLT Survey Telescope (VST; \citealt{VST}). 
SUDARE is a project that aims at analyzing the trend of the rates of different supernova (SN) types in the redshift range $0.3 < z < 0.8$. In addition, it investigates possible correlations with the properties of the host galaxies, and any dependence of such correlations on redshift and/or on the stellar population the SNe belong to. The survey concerns two distinct, well-known sky areas, the above-mentioned COSMOS field and CDFS. Both regions have been widely surveyed by a large contingent of ground- and space-based observatories: this makes them an ideal testing ground, given the wealth of multiwavelength data, from X-rays to radio, available for the validation of the method. 

 The SUDARE survey covers 1 sq. deg. for COSMOS and 4 sq. deg. for the CDFS. Observations started in late 2011, and are available in the $g$, $r$, and $i$ bands. In \citet{decicco} we published the results of the analysis of the first five months of observations (27 visits) for the COSMOS field. Soon after, we presented in \citet{falocco} a similar analysis of the first 2 sq. deg. available for the CDFS: in this case we had 27 visits for one square degree and 22 visits for the other one, spanning five and three months, respectively. We are currently investigating the other half of the surveyed CDFS area, and will present the corresponding results in Poulain et al. (in prep.).
 
 The present paper is dedicated to the analysis of the VST-COSMOS data. The COSMOS campaign has now been extended to a $> 3$ yr baseline, doubling the number of observing visits. Here we exploit the longer baseline and increased number of observations, in order to verify the predictions of our previous works and increase the number of detected AGNs. Also, we better constrain the effectiveness of variability selection with respect to other methods, in order to predict the performance of future surveys. We retrieve a larger sample of optically variable AGN candidates due to their typical red-noise variability, thus making progress toward a more complete and effective census of the AGN population in the field. 
 
The present analysis, as well as all of our works published so far, focuses on $r$-band data for both fields. Additional works investigating the other bands, and a combination of all of them, are currently in preparation (see Section \ref{section:discussion}).

The paper is organized as follows: in Section \ref{section:vst} we describe our dataset; Section \ref{section:selection} illustrates the steps to obtain source catalogs from the various visits, and the reduction process leading to the selection of a robust sample of optically variable AGN candidates. Section \ref{section:analysis} deals with the multiwavelength properties of the selected sample, and describes the various diagnostics used to validate the nature of our AGN candidates. In Section \ref{section:discussion} we gather together our results and discuss our findings, also comparing them to the ones in \citet{decicco}.

\section{The VST-COSMOS data}
\label{section:vst}
The VST, located at Cerro Paranal Observatory, is a 2.65\,m optical telescope with a 0.938\,m-diameter secondary mirror, a modified Ritchey-Chreti\'{e}n configuration, and an alt-azimuth mount. Its detector, OmegaCAM \citep{kuijken}, is a mosaic of 32 CCDs, corresponding to a total of 268 million 15 $\mu$m-size pixels over a $26\times26$ cm$^2$ area. The focal plane scale is $0\farcs214$/pixel, and the corresponding field of view (FoV) is $1^\circ\times1^\circ$.
The VST-COSMOS observations consist of three observing seasons (hereafter, seasons), covering the span from December 2011 to March 2015, including the data from the season already studied in \citet{decicco}.

Each visit\footnote{Throughout the present work we will make a distinction between visits and exposures, a visit being the combination of a number of exposures corresponding to the same observing block (OB). We note that in \citet{decicco} we used the word ``epoch'', instead of ``visit''.} consists of several (usually five, but see Table \ref{tab:dataset} for details) dithered exposures corresponding to individual $1^\circ\times1^\circ$ pointings. Exposure reduction and combination were performed by means of the VST-Tube pipeline \citep{grado}, designed to process VST data. An overview of the processing steps is reported in \citet{decicco}, while a more exhaustive description can be found in \citet{capaccioli15}. Essentially, the pipeline takes care of overscan correction, bias subtraction, and flat-field correction, together with CCD gain harmonization and illumination correction; astrometric correction and photometric calibration follow, then single exposures are combined together into visits. 

The full dataset originally consisted of 65 visits, but we excluded 11 visits from the present analysis, as detailed below; information about the remaining 54 visits, which constitute our final dataset, is reported in Table \ref{tab:dataset}. The first 26 visits in the table cover a five-month baseline and were used for our previous work, described in \citet{decicco}. 

A weight map corresponding to each visit accounts for the different noise level of each pixel by associating them a weight, defined as the reciprocal of the pixel variance. As reference, we used the same deep stacked image as in \citet{decicco}, produced as the median of all the exposures having a seeing full width at half maximum (FWHM) $< 0\farcs80$; the corresponding exposure time is 19800 s, and the limiting magnitude is $r(\mbox{AB})\approx26$ mag at $\sim5\sigma$ above the background r.m.s., while single visits are generally characterized by $r(\mbox{AB}) \lesssim 24.6$ mag for point sources, at the same confidence level\footnote{The depth of VST images is within one magnitude of what LSST is expected to deliver ($r \approx 24.7$ mag for a single DDF visit; see \citealt{ddf}).}. Although produced only from the first five months of our observing campaign, this stack is obviously deeper than any of our individual visits, and sufficient for use as a static reference image.

In the present work we focus on $r$-band data only, which have a three-day observing cadence, while for the $g$ and $i$ bands the cadence is $\sim10$ days; in each case there are several gaps, depending on a number of observational constraints. In the following, magnitudes are quoted in the AB system, unless otherwise stated.

\begin{table*}[htb]
\caption{\footnotesize{COSMOS dataset. Visit number, OB identification number, date, and seeing FWHM for the 54 visits used for the present analysis. Visits are listed in chronological order, and the sequence of their IDs in the first column lacks some numbers because these correspond to the eight visits that we excluded from the analysis (see Section \ref{section:aperture}). Visit 53 was obtained by the combination of ten exposures, for a total exposure time of 3600 s; for all the remaining visits, five exposures were combined together, and the total exposure time is 1800 s.}} 
\label{tab:dataset}      
\begin{minipage}{0.5\textwidth}
\begin{tabular}{c c c c}
\toprule
visit & OB-ID & obs. date & seeing (FWHM)\\
 & & & (arcsec)\\
\midrule
1 & $\mbox{611279}$ & 2011-Dec-18 & $0.64 $\\
2 & $\mbox{611283}$ & 2011-Dec-22 & $0.94 $\\
3 & $\mbox{611287}$ & 2011-Dec-27 & $1.04 $\\
4 & $\mbox{611291}$ & 2011-Dec-31 & $1.15 $\\
5 & $\mbox{611295}$ & 2012-Jan-02 & $0.67 $\\
6 & $\mbox{611299}$ & 2012-Jan-06 & $0.58 $\\ %
7 & $\mbox{611311}$ & 2012-Jan-18 & $0.62 $\\
8 & $\mbox{611315}$ & 2012-Jan-20 & $0.88 $\\	
9 & $\mbox{611319}$ & 2012-Jan-22 & $0.81 $\\	
10 & $\mbox{611323}$ & 2012-Jan-24 & $0.67 $\\
11 & $\mbox{611327}$ & 2012-Jan-27 & $0.98 $\\	
12 & $\mbox{611331}$ & 2012-Jan-29 & $0.86 $\\	
13 & $\mbox{611335}$ & 2012-Feb-02 & $0.86 $\\	
14 & $\mbox{611351}$ & 2012-Feb-16 & $0.50 $\\
15 & $\mbox{611355}$ & 2012-Feb-19 & $0.99 $\\	
16 & $\mbox{611359}$ & 2012-Feb-21 & $0.79 $\\	
17 & $\mbox{611363}$ & 2012-Feb-23 & $0.73 $\\	
18 & $\mbox{611367}$ & 2012-Feb-26 & $0.83 $\\	
19 & $\mbox{611371}$ & 2012-Feb-29 & $0.90 $\\	
20 & $\mbox{611375}$ & 2012-Mar-03 & $0.97 $\\	
21 & $\mbox{611387}$ & 2012-Mar-13 & $0.70 $\\	
22 & $\mbox{611391}$ & 2012-Mar-15 & $1.08 $\\	
23 & $\mbox{611395}$ & 2012-Mar-17 & $0.91 $\\	
24 & $\mbox{768813}$ & 2012-May-08 & $0.74 $\\	
25 & $\mbox{768817}$ & 2012-May-11 & $0.85 $\\	
26 & $\mbox{768820}$ & 2012-May-17 & $0.77 $\\	
28 & $\mbox{986611}$ & 2013-Dec-27 & $0.72 $\\	
    & 				    &			    &	 	  \\
\bottomrule
\end{tabular}
\end{minipage} \hfill
\begin{minipage}{0.5\textwidth}
\begin{tabular}{c c c c}
\toprule
visit & OB-ID & obs. date & seeing (FWHM)\\
 & & & (arcsec)\\
\midrule
29 & $\mbox{986614}$ & 2013-Dec-30 & $1.00 $\\	
30 & $\mbox{986617}$ & 2014-Jan-03 & $0.86 $\\	
31 & $\mbox{986620}$ & 2014-Jan-05 & $0.81 $\\	
32 & $\mbox{986626}$ & 2014-Jan-12 & $0.73 $\\	
33 & $\mbox{986630}$ & 2014-Jan-21 & $1.18 $\\
34 & $\mbox{986633}$ & 2014-Jan-24 & $0.80 $\\	
37 & $\mbox{986648}$ & 2014-Feb-09 & $1.28 $\\
38 & $\mbox{986652}$ & 2014-Feb-19 & $0.89 $\\	
39 & $\mbox{986655}$ & 2014-Feb-21 & $0.93 $\\	
40 & $\mbox{986658}$ & 2014-Feb-23 & $0.81 $\\	
41 & $\mbox{986661}$ & 2014-Feb-26 & $0.81 $\\	
42 & $\mbox{986664}$ & 2014-Feb-28 & $0.77 $\\	
44 & $\mbox{986670}$ & 2014-Mar-08 & $0.91 $\\	
45 & $\mbox{986674}$ & 2014-Mar-21 & $0.96 $\\	
46 & $\mbox{986677}$ & 2014-Mar-23 & $0.92 $\\	
47 & $\mbox{986680}$ & 2014-Mar-25 & $0.66 $\\
48 & $\mbox{1095777}$ & 2014-Mar-29 & $0.89 $\\	
49 & $\mbox{1095783}$ & 2014-Apr-04 & $0.58 $\\	
50 & $\mbox{986683}$ & 2014-Apr-07 & $0.61 $\\	
51 & $\mbox{1136410}$ & 2014-Dec-03 & $1.00 $\\	
53 & $\mbox{1136457}$ & 2015-Jan-10 & $0.71 $\\	
55 & $\mbox{1136481}$ & 2015-Jan-28 & $0.90 $\\	
56 & $\mbox{1136490}$ & 2015-Jan-31 & $0.73 $\\	
57 & $\mbox{1136503}$ & 2015-Feb-15 & $0.70 $\\	
60 & $\mbox{1136531}$ & 2015-Mar-10 & $0.80 $\\	
61 & $\mbox{1136540}$ & 2015-Mar-14 & $0.84 $\\	
62 & $\mbox{1136543}$ & 2015-Mar-19 & $1.00 $\\	
stacked & - & - & $0.67 $\\
\bottomrule
\end{tabular}
\end{minipage} \hfill
\end{table*}

\section{Selection of variable sources}
\label{section:selection}
As a first step, we visually inspected each visit in our dataset in order to verify its quality: this led to the exclusion of three visits because of severe aesthetic artifacts and defects\footnote{The defects in one of the three visits are mainly arcs originating from reflections internal to the telescope, due to the presence of nearby bright stars, while the other two excluded visits are characterized by unusually high noise: this was caused by a very strong and inhomogeneous illumination, originating from moonlight reflections by clouds and some tracking problems with the telescope.}. 

The method we adopt to identify optically variable sources, and hence define our sample of AGN candidates, follows the approach proposed by \citet{Trevese}.

\subsection{Catalog production and aperture selection}
\label{section:aperture}
A catalog of sources was obtained from each visit making use of \emph{SExtractor} \citep{bertin}: this contains a number of parameters for each object, including positional coordinates, half-light radii, and magnitudes through a set of fixed apertures. We fed the above-mentioned weight maps to \emph{SExtractor} so that, in the process of catalog extraction, the different quality of each pixel in each visit is taken into account.

AGN identification requires an aperture size allowing the collection of the bulk of the flux from the nucleus of a galaxy, while minimizing the contribution from the host galaxy itself and possible nearby sources. 
In \citet{decicco} we measured the source flux within a 2\arcsec-diameter, which typically encloses $\approx70\%$ of the flux from a point-like object, and then, in order to take into account the effect of seeing, we computed corrective factors for each visit making use of growth curves of reference stars. A growth curve shows how the fraction of flux collected from a source changes as a function of the aperture size\footnote{In \citet{decicco} for each visit we determined a corrective factor: we defined it as the ratio of the flux from the reference star enclosed in a 2\arcsec-diameter aperture to the flux corresponding to 90\% of the total. In this way, independent of the visit seeing, each corrected magnitude corresponds to 90\% of the flux collected in the chosen aperture.We note that the 90\% choice is arbitrary: what matters is that the fraction of collected flux be the same for each source in each visit. The reference stars we chose were detected in each visit and were not saturated; they did not have close neighbors, and were distant from possibly defected regions of the image (see next section).}.

\begin{figure}[tb]
\centering
            {\includegraphics[width=8.9cm]{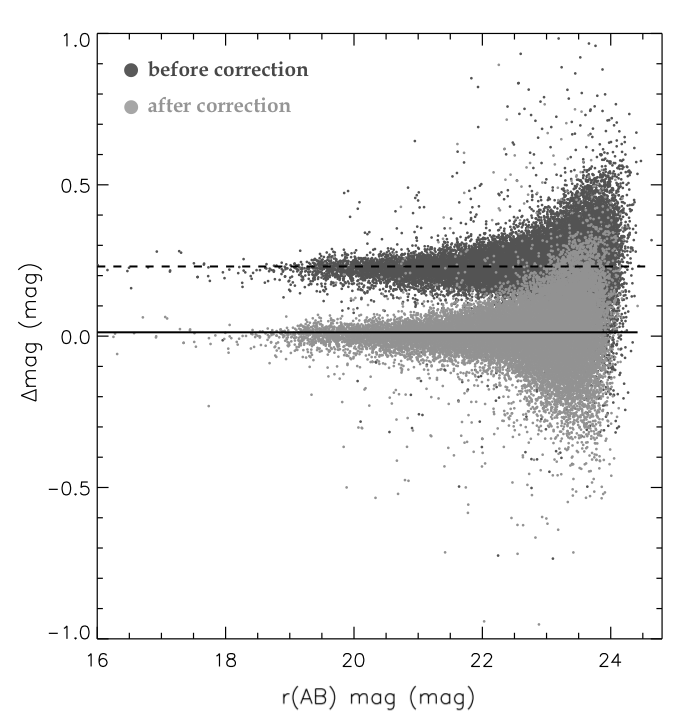}}
\caption{\footnotesize Magnitude difference between visit 3 and visit 49, which is used as a reference. Though all the sources with an available measurement of their magnitude are shown, the ones used to determine the aperture corrective factors have \emph{r}(AB) magnitudes in the range 16--21. Dark and light gray dots represent magnitudes before and after the correction, respectively; similarly, the dashed and solid lines correspond to the average magnitude difference before and after the correction, respectively. The first defines the corrective factor which, in this case, is $\approx 0.22$. The obtained value varies depending on the visit seeing, being larger for higher seeing values; visit 3 has a seeing value of 1\farcs04, which is one of the largest in our set of visits.}\label{fig:delta_mag}
\end{figure}

Here we adopt a different approach, following \citet{Trevese}, which essentially consists of normalizing all visits to a reference one. In our case we chose visit 49, which has the second-best seeing (see Table \ref{tab:dataset}) and is free from significant aesthetic artifacts \footnote{Visit 14, which is the one with the best seeing, is affected by bad pixels and artifacts.}. Then, for each visit, we selected all the sources with 2\arcsec-diameter aperture \emph{r}(AB) magnitudes in the range 16--21 (so as to avoid very bright/saturated or faint and noisy objects) and computed the average magnitude difference $\langle\Delta\mbox{\tiny{mag}}\rangle$ with respect to the reference visit. The catalogs of sources obtained from different visits were matched requiring the distance between positional coordinates to be $< 1\arcsec$. We subtracted from all 2\arcsec-aperture magnitudes in each visit the corresponding corrective factor in order to account for seeing and calibration differences. The magnitude difference between visits 3 and 49, for all the sources in the field, before and after the correction, is shown in Fig. \ref{fig:delta_mag} as an example. 

\begin{figure}[tb]
\centering
            {\includegraphics[width=\textwidth/2]{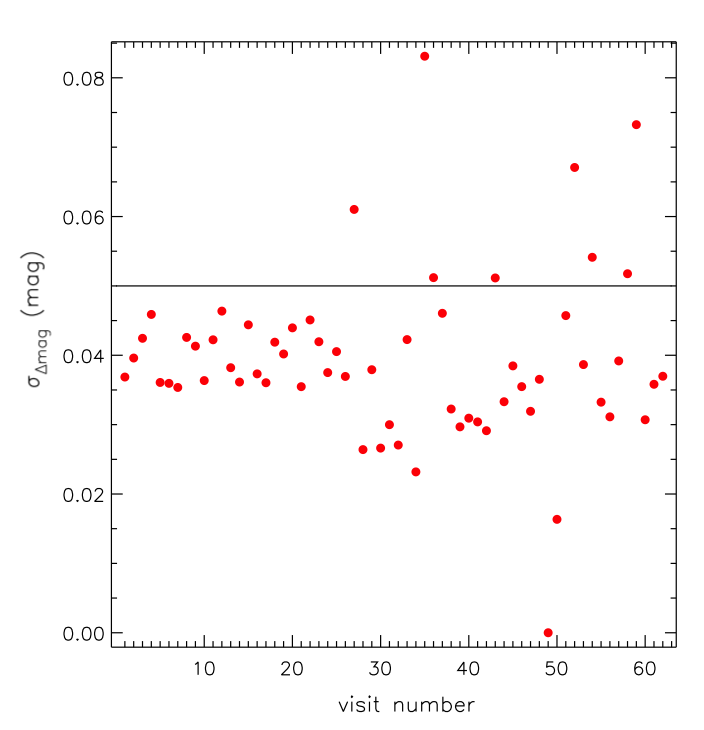}}
\caption{\footnotesize $\sigma_{\Delta\mbox{\tiny{mag}}}$ for each visit. The sources used to determine the aperture corrective factors have \emph{r}(AB) magnitudes in the range 16--21. The points above the black line correspond to the eight visits that we excluded from our dataset.}s\label{fig:epoch_selection}
\end{figure}

As a further step, we computed the r.m.s. deviation $\sigma_{\Delta\mbox{\tiny{mag}}}$ of $\Delta\mbox{\tiny{mag}}$ for each visit, in order to quantify the calibration uncertainty with respect to the reference visit. $\sigma_{\Delta\mbox{\tiny{mag}}}$ for each visit is reported in Fig. \ref{fig:epoch_selection}. We decided to exclude from our dataset the visits characterized by $\sigma_{\Delta\mbox{\tiny{mag}}}$ values higher than 0.05; these correspond to eight visits, represented by the most scattered points in the figure\footnote{Scattered points corresponding to low values are not to be taken into account, as their photometry is the closest to the one in the reference visit.}. The adopted limit is arbitrary and indeed there are some excluded visits very close to the threshold. 
We tested how the inclusion of these four visits would affect the analysis. Although their inclusion would add a few ($< 10$) more sources to the sample of optically variable AGN candidates, it would also introduce several dozens of contaminants. Thus, although excluding these visits may be considered a conservative approach,
their exclusion did not significantly impact our results.

\subsection{Defects and masks}
\label{section:masks}
Observations of the COSMOS field from the first season correspond to the very beginning of VST activity, and various early visits turned out to be affected by a number of defects in the detector electronics, as well as aesthetic artifacts; VST data-users, including us, had to address them for the first time, and the most natural path to follow resulted in a conservative approach. Some of the problems arose from a poor knowledge of the detector response, and were fixed in the following months. In particular, the most relevant problem was a CCD characterized by random variations in its gain factor, which resulted in a large concentration of fake sources in the corresponding area of the detector; this led to the exclusion of the image region corresponding to that CCD from the affected visits. Similarly, other regions were excluded in some or all the visits, depending on the circumstances. Examples are the regions at the edge of each visit (corresponding to a $\approx 2\arcmin$--$4\arcmin$ width each side), characterized by a very low signal-to-noise ratio (S/N), satellite tracks, regions affected by reflections of scattered light, and --last but not least-- bright star halos. The last ones are a common feature in wide-field surveys, and constitute the main problem that we had to address by applying suitable masks to the images at issue. Bright stellar halos and spikes were masked making use of the regions produced by the {\sl Pulecenella} code developed by \citet{huang}. These initial masks were improved upon by adding, when necessary, additional eye-selected regions to be excluded, for each visit.
 
Specifically, for one individual visit, an additional region affected by reflections of scattered light was masked. Satellite tracks were not masked, but the potential problems they could cause were minimized by resorting to a $\sigma$-clipping algorithm (see Section \ref{section:sample}).
The masking process reduced the number of objects in each catalog by a couple tens of thousands, representing 20\%--25\% of the whole source catalog depending on the seeing value (higher seeing required larger masked areas).

\subsection{The sample of AGN candidates}
\label{section:sample}
We cross-matched the catalogs of sources obtained for each visit by matching positional coordinates within a $1\arcsec$ radius, in order to get the light curves of all the detected objects. Source separations are $<$ 0\farcs26 in 95\% of the cases. The master catalog thus obtained includes all the variable and non-variable sources detected in at least two visits and with an average magnitude \emph{r}(AB) $\leq 23.5$ mag. Our previous work focused on sources with a \emph{r}(AB) $\leq 23$ mag; here we extend the sample down to fainter magnitudes, consistent with the single-visit completeness limit.

The master catalog contains 25452 sources, including a number of objects detected in just a few visits, many of these are either fast transients and/or more likely spurious sources. In order to minimize contamination from such objects, we make a further cut and require the sources in our final catalog to be detected in at least half the visits.

We note that we adopt a different criterion than in \citet{decicco}, where we required sources to be detected in at least 20\% of the visits. The minimum number of visits is somewhat arbitrary, as it depends on the type of source to detect/exclude and its variability. Here we increased the minimum visit requirement based on our experience with the first five months of data, and considering the longer time span of our light curves (which increases the likelihood of detecting variability from fainter AGNs). We find that anyway 90\% of the sources in our master catalog are detected in at least half the visits. This percentage would rise to 99\% if we required sources to be detected in 20\% of the visits, but the sample of variable AGN candidates would be almost the same as the one we obtain with the adopted threshold, with a couple dozens additional sources that would be mostly contaminants. We therefore opted for the more conservative threshold.

The number of visits in each light curve also affects the detection significance for variable sources due to the different number of degrees of freedom in each light curve \citep[see discussion in][]{decicco}. To assess the relevance of this effect we performed simulations including each time a different number of visits, and we found that the detection threshold for variable sources differs by less than 15\% between sources detected in all visits and those detected in just half of them.

The applied selection criteria returned a sample of 22927 sources (hereafter, the \emph{main sample}); we point out that this includes both variable and non-variable sources.

In order to define a variability threshold, we computed the average magnitude $\langle{\mbox{mag}}_i^{lc}\rangle$ and the corresponding r.m.s. deviation $\sigma_i^{lc}$ for each source $i$ from the corresponding light curve:
\begin{equation} 
\small
\langle{\mbox{mag}}_i^{lc}\rangle = \frac{1}{N_{vis}}\sum_{j=1}^{N_{vis}}\mbox{mag}_i^j\mbox{   ,}\qquad\sigma_i^{lc}={\left[\frac{1}{N_{vis}}\sum_{j=1}^{N_{vis}}{(\mbox{mag}_i^j-\langle{\mbox{mag}}_i^{lc}}\rangle)^2\right]}^{\frac{1}{2}}\mbox{,}\label{eqn:avg_stdev}
\end{equation}
where the superscript $lc$ stands for ``light curve'', $j$ runs through the visits, and $N_{vis}$ is the number of visits where we detect the source $i$. 
Both $\langle{\mbox{mag}}^{lc}\rangle$ and $\sigma^{lc}$ were obtained through a $\sigma$-clipping algorithm, rejecting $> 5\sigma$ outliers, in order to minimize spurious contributions to magnitude variations due, for instance, to residual aesthetic defects (e.g satellite tracks, cosmic rays, stellar diffraction spikes, etc.). 

\begin{figure}[tb]
\centering
            {\includegraphics[width=9cm]{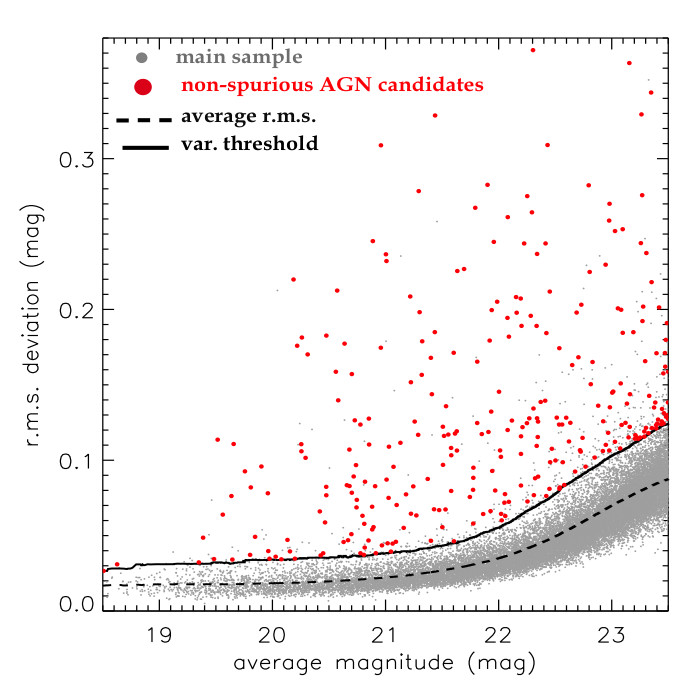}}
\caption{\footnotesize Light curve r.m.s. $\sigma^{lc}$ as a function of the average magnitude $\langle{\mbox{mag}}^{lc}\rangle$ for all the sources in the \emph{main sample} (small gray dots). The dashed line represents the running average of the r.m.s., while the solid line defines the variability threshold. Objects above the threshold are considered to be variable, and large red dots identify non-spurious AGN candidates (flag 1 or 2, see further in the text).}\label{fig:stdev_vs_mag}
\end{figure}

Once we measured the properties of all light curves in our \emph{main sample}, we extracted the sample of variable sources. We initially define as variable candidates all sources exhibiting a $\sigma^{lc}$ in excess of the 95th percentile of the $\sigma^{lc}$ distribution, over a running 0.5 mag-wide bin centered on the magnitude of each source. 
We identified 482 sources ($2.1\%$ of the \emph{main sample}) above the variability threshold, which constitute our preliminary sample of AGN candidates. Figure \ref{fig:stdev_vs_mag} shows the variability threshold and the sample of AGN candidates in the plane of $\sigma^{lc}$ vs. $\langle{\mbox{mag}}^{lc}\rangle$.

We expect the variability of some sources in our preliminary sample to be spurious due to a number of factors: contamination from a close companion\footnote{In principle, a change in the input extraction parameters could help reduce such occurrences but, in any case, when a source has a very close companion and, in particular, when this object is very bright, it is not possible to tell whether variability is an intrinsic property of the source or arises from visit-to-visit point spread function (PSF) variations combined with contamination from the companion.}; the irregular morphology of some galaxies, which makes it difficult to identify the source centroid in the visits with the worst seeing values; the presence of problematic areas of the detector (e.g., hot pixels) or noise-dominated regions (e.g., the edges of a frame) where sources happen to fall. We also expect our sample to include a number of contaminants, such as variable stars and transient events (e.g., SNe). The identification and analysis of SNe in our dataset is currently ongoing (Ragosta et al., in prep.); in \citet{decicco} we measured a $14\%$ contamination rate due to SNe in our final sample of variable sources\footnote{The contamination is defined as the number of confirmed non-AGNs divided by the number of AGN candidates; it can also be computed for specific classes of sources (e.g., SNe, stars) when information about them is available.}.

Our preliminary sample of optically variable sources includes 90 objects that suffer from blending with neighboring galaxy at least in the visits with worse seeing. We inspected the snapshots of each source in each visit, and they revealed that the centroid position can vary substantially, migrating from one object to the other, depending on the visit. This means that, in either case, the corresponding measured flux is possibly wrong, hence we excluded them from our analysis. This leaves 392 candidates. In order to identify and reject spurious sources and contaminants, we visually inspected the snapshots of each source in each visit, and flagged sources with a quality label ranging from 1 to 3 on the basis of the following guidelines:
\begin{enumerate}
\item strong candidate; no problems or defects detected (232 sources);
\item likely variable candidate; neighbor potentially affecting the source, or minor problems detected (67 sources); 
\item very doubtful variability; likely spurious (93 sources).
\end{enumerate}
Sources are flagged as 3 either because of their irregular shape (43 objects), which does not allow reliable identification of a center, or because of the presence of a nearby saturated object (50 objects); in this case we used the same empirical selection criterion that we set in our previous work, and required the centroid-to-centroid distance to be $\leq$ 2\arcsec$\mbox{ }$and the magnitude difference of the two sources to be $< 1.5$ mag for a source to be a flag 3, as we assumed that, in such a case, the light from one source strongly affects the other. 

We point out that the sample cleaning process can be automated in part using \emph{SExtractor} flags and other diagnostics to exclude \emph{a priori} sources with close neighbors, those with cross-contamination, and so on. In this case we resorted to a manual approach due to the limited number of candidates to inspect and to have a reference sample to optimize the automatic procedure in the future; the implementation is already in progress for our already-mentioned follow-up work combining $g$, $r$, and $i$ light curves, and will be an unavoidable approach when dealing with any future larger sample. 

\begin{figure}[h!]
 \centering
   {\includegraphics[width=9cm]{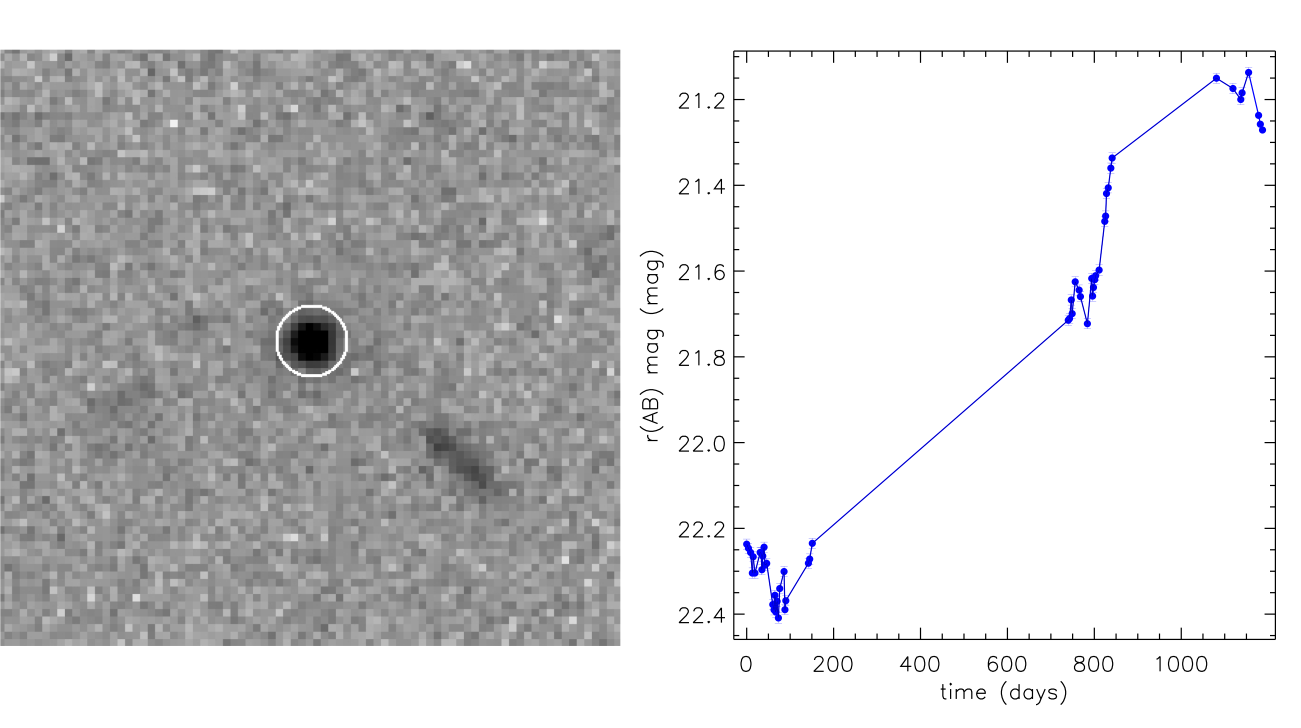}}
   {\includegraphics[width=9cm]{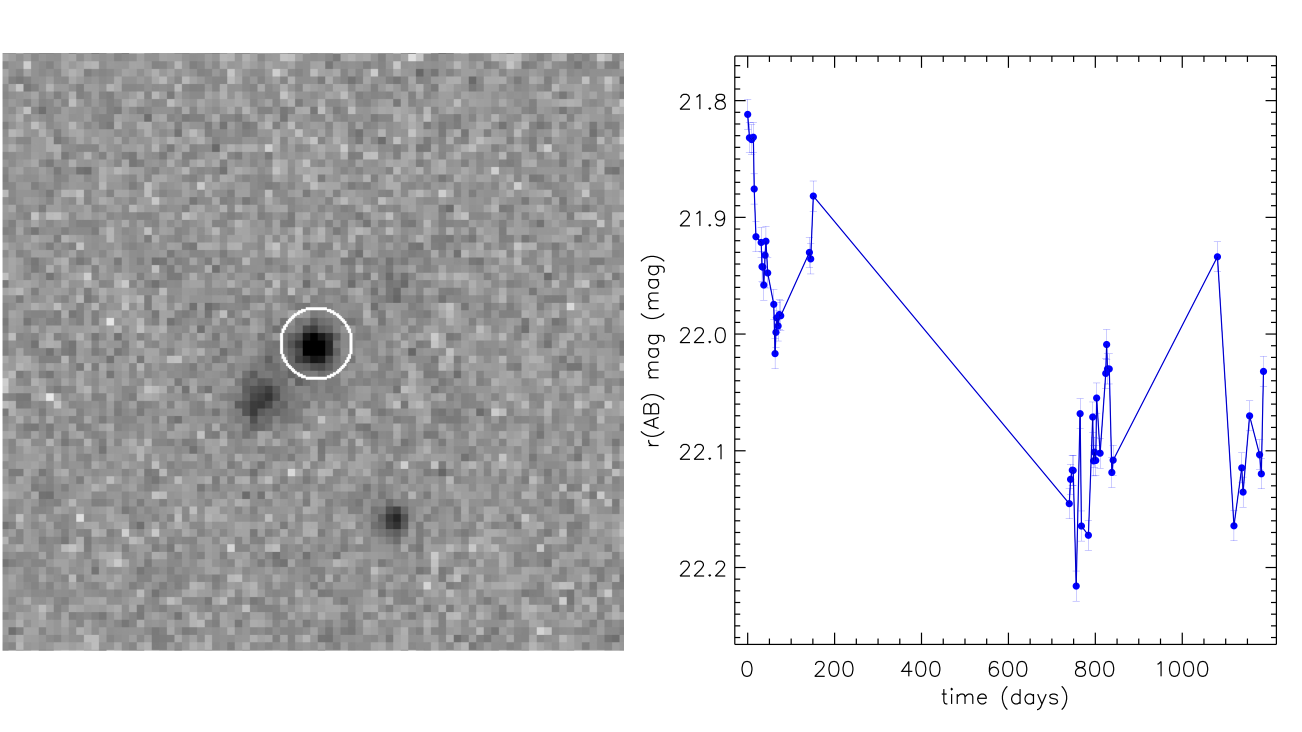}}  
   {\includegraphics[width=9cm]{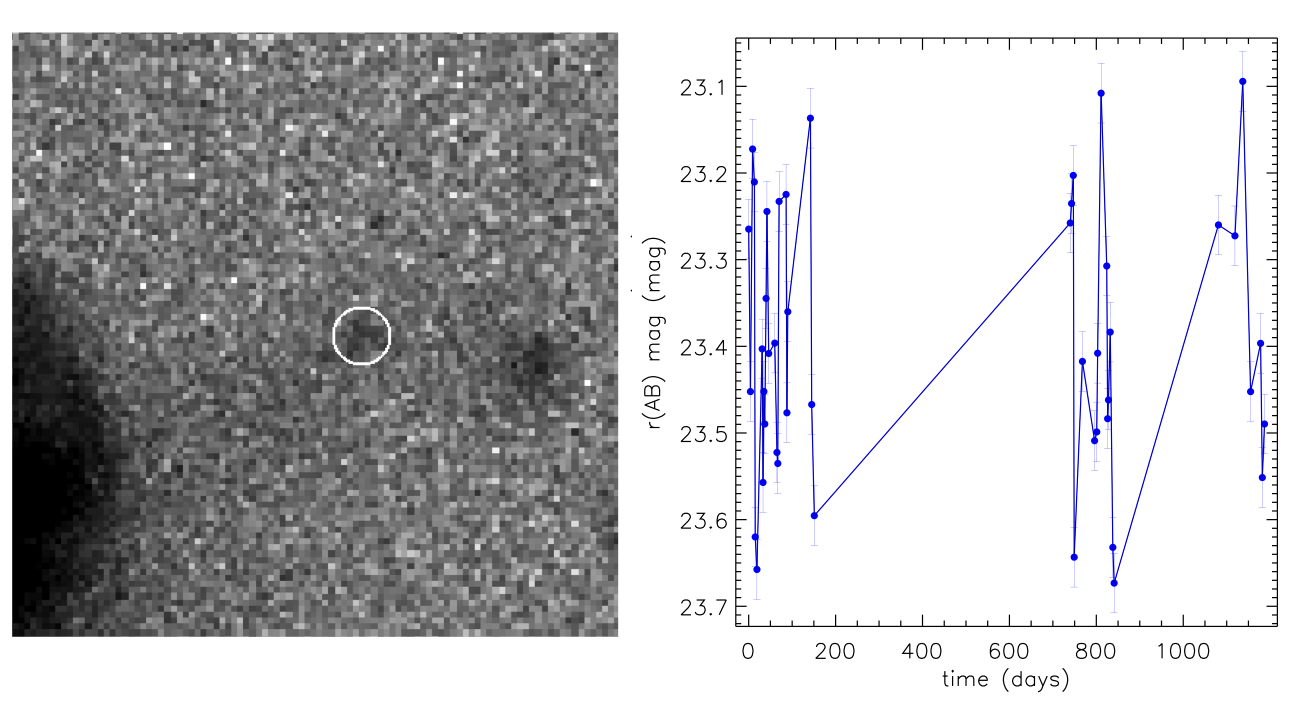}}
 \caption{\footnotesize{Examples of variable AGN candidates assigned to different quality classes, with corresponding light curves. The images in the upper and middle panel are from visit 49, which has the second-best seeing; the image in the lower panel is from visit 33, which has the worst seeing. White circles correspond to the $2\arcsec$-diameter aperture and are centered on the average object coordinates. Objects labeled 1 (\emph{upper panel}) are generally isolated and free from aesthetic defects. In the case of the objects belonging to class 2 (\emph{middle panel}), potential problems (e.g., the presence of a neighbor) must be taken into account. Objects labeled 3 (\emph{lower panel}) are probably spurious variable sources; the case of a source lying in the halo of an extended and bright neighbor (on the far left) is shown as an example: from the corresponding light curve, we can see that its erratic variability, well in excess of the photometric error, is likely due to an improper subtraction of the  variable background due to the bright neighbor. The error bars for each source are defined as  the 95\% uncertainty on the source magnitude.}}\label{fig:flag_examples}
\end{figure}

In Fig. \ref{fig:flag_examples} we present one source per class as an example; images for flags 1 and 2 are from the visit with the second-best seeing, while for flag 3 we show an image from the worst-seeing visit, in order to give an idea of how hard it is to identify the source above the complex background due to the bright extended neighbor. The variability analysis that we describe in the following is limited to the sources labeled 1 or 2 (hereafter, the \emph{robust sample}), consisting of 299 sources and hence constituting $1.3\%$ of the \emph{main sample} and 62\% of the initial AGN candidates with \emph{r}(AB) mag $\leq 23.5$ mag.  

\section{Validation and characterization of the AGN candidates}
\label{section:analysis}
In the present section we investigate the nature and properties of the sources in our \emph{robust sample}, aiming at confirming AGNs and at classifying as many of the remaining sources as possible. We intend to assess to what extent optical variability is a reliable indicator of the presence of an AGN, and hence a powerful alternative to other more expensive and/or time-consuming techniques (e.g., X-ray identification) usually employed in the search for AGNs.

Due to the widespread literature about the COSMOS field, several catalogs at different wavelengths are available for most of the sources, and classification already exists for part of our sample. This allowed us to validate our results by means of extensive ancillary data.

\subsection{X-ray counterparts}
\label{section:X}
X-ray emission, especially if coupled with variability, is the strongest and most reliable indicator of the presence of an AGN \citep[e.g.,][]{brandt&alexander}. In the present work we made use of two X-ray catalogs of COSMOS sources to identify possible X-ray counterparts for our \emph{main sample}:
\begin{itemize}
\item[--] the catalog containing optical and near-infrared (NIR) counterparts of the sources in the \emph{Chandra}-COSMOS Legacy Catalog \citep{marchesi, civano16}, consisting of 4016 X-ray emitters over a 2.2 sq. deg. area; the catalog is the result of a 4.6 Ms program with an exposure of $\simeq160$ ks in the central 1.5 sq. deg. and $\simeq 80$ ks in the surrounding area. The limiting depth corresponds to fluxes of $2.2\times10^{-16}$ erg\,cm$^{-2}$ s$^{-1}$,  $1.5\times10^{-15}$ erg\,cm$^{-2}$ s$^{-1}$, and $8.9\times10^{-16}$ erg\,cm$^{-2}$ s$^{-1}$ in the 0.5--2, 2--10, and 5--10 keV bands, respectively. The catalog provides a considerable volume of information, including optical counterparts for 3899 ($97\%$) of the 4016 sources, spectroscopic classification (broad-line AGN, i.e., BLAGN; non-BLAGN; star) for $\approx42\%$ of the sample, and photometric classification (obscured/unobscured AGN; galaxy; star) based on the fitting of the spectral energy distribution (SED) for $96\%$ of the sample;
\item[--] the \emph{XMM}-COSMOS Point-like Source catalog \citep{brusa}, including 1674 X-ray sources with optical counterparts. The corresponding observations ($\simeq 60$ ks) in this case are shallower and a little less extended (2 sq. deg. area) than the \emph{Chandra}-COSMOS Legacy Survey; the flux limits in the 0.5--2\, keV, 2--10\, keV, and 5--10\, keV energy bands are, respectively, $\approx1.7\times10^{-15}$ erg\,cm$^{-2}$ s$^{-1}$, $\approx9.3\times10^{-15}$ erg\,cm$^{-2}$ s$^{-1}$, and $\approx1.3\times10^{-14}$ erg\,cm$^{-2}$ s$^{-1}$ over $90\%$ of the area. Spectroscopic classification (BLAGN; narrow-line AGN, i.e., NLAGN; normal/star-forming galaxy) is available for about half the sample, and a best-fit SED template by \citet{salvato} is provided for $97\%$ of the sample. Though this catalog is far shallower than the \emph{Chandra} catalog, it provides some useful details about the nature of the sources (see Section \ref{section:spec_class}).  
\end{itemize}

The two X-ray catalogs provide information for 1815 X-ray sources in the VST-COSMOS FoV (excluding masked areas), with coordinates for their optical counterparts reported in the X-ray catalogs themselves. Nevertheless, not all of them have a VST counterpart. Specifically, we found:
\begin{itemize}
\item[--] 719 X-ray sources with magnitude $r$(AB) $< 23.5$ mag and a VST counterpart;
\item[--] 575 sources with magnitude $r$(AB) $> 23.5$ mag\footnote{Magnitude values where retrieved either from the COSMOS2015 catalog presented in \citet{laigle}, or from the \emph{XMM}-COSMOS catalog; the latter reports $r$(AB) magnitudes from \citet{capak} for 98\% of the sources. In only two cases magnitudes are from the COSMOS Intermediate and Broad Band Photometry catalog.}, hence not taken into account in what follows;
\item[--] 512 sources with a VST counterpart but detected only in a few visits since they are close to the detection limit, and hence excluded from the \emph{main sample}, according to the visit threshold we set (see Section \ref{section:sample}); 
\item[--]five sources with $r$(AB) $< 23.5$ mag and with a very close, very bright neighbor, which prevented the detection in VST data;
\item[--]three sources for which we find a VST counterpart with a matching radius larger than $1\arcsec$;
\item[--]one source which appears very blurred in the VST images, and with no magnitude estimate in either of the two X-ray catalogs. 
\end{itemize}

The sample of X-ray sources that have a VST counterpart and fulfill our selection criteria hence consists of 719 objects (hereafter, the \emph{X-ray sample}), or $3\%$ of the 22927 sources in the \emph{main sample}. The match of the \emph{robust sample} with the \emph{X-ray sample} revealed 250 AGN candidates (83.6\% of the \emph{robust sample}: hereafter, X-ray emitting variable AGN candidates) with an X-ray counterpart; this implies that 250 out of 719 X-ray sources with bright optical counterparts (35\%) appear optically variable in our catalog. 

All but one of the 250 X-ray emitting variable AGN candidates have an X-ray counterpart in the deeper \emph{Chandra} catalog, and we thus preferentially quote the \emph{Chandra} values. 

\begin{figure}[tb]
 \center
   {\includegraphics[width=9cm]{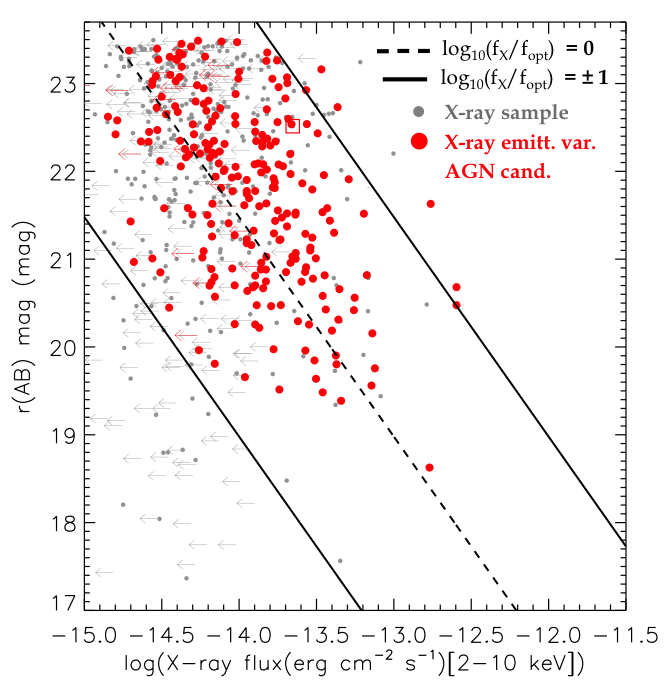}}
\caption{\footnotesize \emph{r}(AB) magnitude vs. hard (2--10\,keV) X-ray flux for the 250 X-ray emitting variable AGN candidates (large red symbols denote \emph{Chandra} detections, open box denotes \emph{XMM} detection). The small grey symbols denote the remainder of the \emph{X-ray sample}. Leftward arrows indicate that only upper limits of the X-ray flux values are available. The dashed line corresponds to X/O $=0$; the lower and upper solid lines represent X/O $=-1$ and X/O $=1$, respectively, and define the AGN locus.}\label{fig:XO}
\end{figure}

The ratio of the X-ray-to-optical flux (X/O) of a source is traditionally defined as \citep{maccacaro}
\begin{equation}
X/O=\log(f_{\scriptscriptstyle{X}}/f_{opt})=\log f_{\scriptscriptstyle{X}} + \frac{\mbox{mag}_{opt}}{2.5} + C \mbox{ ,}\label{eqn:XO}
\end{equation}
where $f_{\scriptscriptstyle{X}}$ is the X-ray flux measured in a chosen energy band, while $\mbox{mag}_{opt}$ is the optical magnitude at a chosen wavelength, and $C$ is a constant which depends on the magnitude system adopted for the observations; in our case, the X-ray fluxes are from the \emph{Chandra} catalog and are measured in the 2--10\,keV band, while the optical magnitudes are the VST \emph{r}(AB) magnitudes; this leads to a value for the constant $C = 1.0752$. Stars and inactive galaxies typically exhibit X/O $< -2$ \citep[e.g.,][]{mainieri, xue}, while AGNs are usually characterized by $-1\le$ X/O $\le 1$, hence the X/O constraints on our sources can help unveil their nature. In Fig. \ref{fig:XO} we show the hard (2--10\,keV) X-ray flux vs. \emph{r}-band magnitude for the sources in the \emph{robust sample} with an X-ray counterpart. It is apparent that all but six of the X-ray emitting variable AGN candidates lie in the AGN locus, while six sources have an X/O $>1$; following \citet{civano}, we consider all 250 X-ray emitting variable AGN candidates as AGNs due to their X/O values\footnote{We note that the sources out of the AGN locus on this diagram could be AGNs as well: indeed, this diagnostic is typically biased against X-ray faint AGNs \citep[see, e.g.,][]{salvato18a}.}. This means that, if no additional information about our \emph{robust sample} were available, on the sole basis of the X/O diagram we could be confident that $83.6\%$ (250/299) are indeed AGNs, representing a lower limit on the \emph{purity} of our sample. We define the \emph{completeness} as the ratio of the sources that we confirm as AGNs (250 objects) and the number of AGNs in our \emph{X-ray sample} (i.e., the X-ray sources that lie in the AGN region on the X/O diagram, namely 668 sources). This results in a fraction of 37\% (250/668) completeness. 
We point out that there is always some degree of uncertainty in the definition of the X/O of a source, because of the intrinsic variability of the source itself combined with the non-simultaneity of the X-ray and optical observations \citep[e.g.,][]{paolillo17, chiaraluce}, and the changing contributions from the host galaxy in the optical band as a function of redshift (see, e.g., Fig. 3 of \citealt{alexander}).

\subsection{Spectroscopic and photometric classification}
\label{section:spec_class}
In Section \ref{section:X} we mentioned that both \emph{Chandra} and \emph{XMM} catalogs provide a spectroscopic and a photometric classification for part of the X-ray emitters in each of the catalogs. In the first one, objects with a spectroscopic classification are labeled as BLAGNs if their spectra show at least one broad (FWHM $> 2000$ km s$^{-1}$) emission line, while non-BLAGNs could be NLAGNs or star-forming galaxies: this is either because most of the sources at issue are characterized by low S/N spectra, or because the waveband in which the spectra are obtained does not allow utilization of optical emission line diagnostics which would help separating the two classes of objects. The photometric classification, instead, is derived through best-fit templates of the broadband SEDs, and sources are divided into unobscured AGNs, obscured AGNs, inactive galaxies, and stars. A cross-match of the two classifications confirmed that $82\%$ of the BLAGNs correspond to unobscured AGNs; the match is not higher because BLAGN SEDs, especially for low-luminosity AGNs, suffer from stellar light contamination; non-BLAGNs are matched to obscured AGNs in $23\%$ of the cases and to galaxies in $74\%$ of the cases \citep{marchesi}. 

The \emph{XMM} catalog classifies sources as BLAGNs, NLAGNs, and inactive galaxies. BLAGNs must fulfill the same criterion as in the \emph{Chandra} catalog; sources flagged as NLAGNs typically have spectra characterized by unresolved high-ionization emission lines with line ratios suggesting AGN activity, while inactive galaxy spectra are generally consistent with those of star-forming or normal galaxies, and have rest-frame hard X-ray luminosity $L_X<2\times10^{42}$ erg\,s$^{-1}$, when detected in the hard X-rays; part of the best-fit SED templates correspond to unobscured (also known as Type 1) and obscured (Type 2) AGNs \citep{brusa}.

In the VST \emph{X-ray sample} (see Section \ref{section:X}) 600 out of 719 sources have a spectroscopic classification from either of the two catalogs; moreover, all but six out of 719 sources have a photometric classification. Among the spectroscopically classified objects, we find 243 unobscured AGNs and 140 obscured AGNs, based on the \emph{Chandra} catalog and always adopting the spectroscopic over photometric classification. While the definitions for unobscured AGNs are the same in the two X-ray catalogs, the label ``non-BLAGN'' in the \emph{Chandra} catalog is ambiguous (see details in Section \ref{section:X}), and hence not sufficient to classify a source. The \emph{robust sample} also includes 11 sources spectroscopically classified as stars.

Our 250 X-ray emitting variable AGN candidates are spectroscopically classified as: 200 unobscured AGNs; 25 obscured AGNs; two stars; five sources with uncertain classification, including two possible inactive galaxies; 18 unclassified sources. Nonetheless, we note that 17 out of the 23 sources not classified as obscured/unobscured AGNs (18 unclassified + 5 uncertain), including the two possible inactive galaxies, have X-ray luminosities $L_X>3 \times10^{42}$ erg\,s$^{-1}$, indicative of AGNs \citep[e.g.,][]{brandt&hasinger}; moreover, by selection all 23 sources lie on the AGN stripe in the X/O diagram. This means that spectroscopic information alone allows to identify 225 AGNs and two stars; combining this information with the X/O information, and taking into account the X-ray luminosities, we can therefore classify as AGNs 248 out of the 250 X-ray emitting variable AGN candidates, excluding the two stars, but including the two possible inactive galaxies.

The completeness of variability selection with respect to the spectroscopically confirmed sample is defined as the ratio of spectroscopically confirmed AGNs (in the \emph{robust sample}) to the expected AGNs (in the \emph{X-ray sample}), and is $59\%$ (225/383) while, if computed separately for unobscured and obscured AGNs, we obtain 200/243 = $82\%$ and 25/140 = $18\%$, respectively. These fractions are much higher than those reported in \citet{decicco}; this is one of the main results of this work and will be discussed further in Section \ref{section:discussion}. 
\subsection{Color-based classification}
\label{section:colors}
Diagrams comparing source colors are widely used to disentangle different classes of objects (see, e.g., \citealt{boutsia, nakos}), based on the fact that sources tend to have distinct SEDs and occupy distinct loci on such diagrams, depending on their nature. 

 In order to obtain color diagrams, we made use of data from two additional COSMOS catalogs:
\begin{itemize}
\item[--] the already-mentioned COSMOS2015 catalog \citep{laigle}, which contains photometry and physical parameters for more than half a million objects in several filters (including intermediate- and narrow-band ones) spanning a wide range of wavelengths, and also provides matches with optical, X-Ray, UV, IR, and radio catalogs, as well as previous versions of the COSMOS multiband catalog. We use this catalog to obtain the magnitudes we need in the $r$, $z$, and $k$ bands, down to a magnitude \emph{r}(AB) $\approx 28$ mag\footnote{As stated in Section \ref{section:sample}, we limit our analysis to sources brighter than 23.5 \emph{r}(AB) mag; we report the magnitude limit of the COSMOS2015 catalog only to show that the magnitude range we investigate here is fully covered.}.

\item[--]the COSMOS ACS catalog \citep{koekemoer, scoville} from the \emph{Hubble Space Telescope} (\emph{HST}), obtained from 575 pointings of the Advanced Camera for Surveys (ACS). This catalog is mainly used because it provides a morphological classification based on the \emph{SExtractor} stellarity index, ranging from 0 (extended source) to 1 (point-like source). Given the exquisite \emph{HST} resolution, this catalog allows us to better distinguish galaxies from stars and quasars, as discussed below.
\end{itemize}

\subsubsection{Optical-NIR diagnostic}
\label{section:rzk}
\begin{figure}[tb]
 \center
   {\includegraphics[width=9cm]{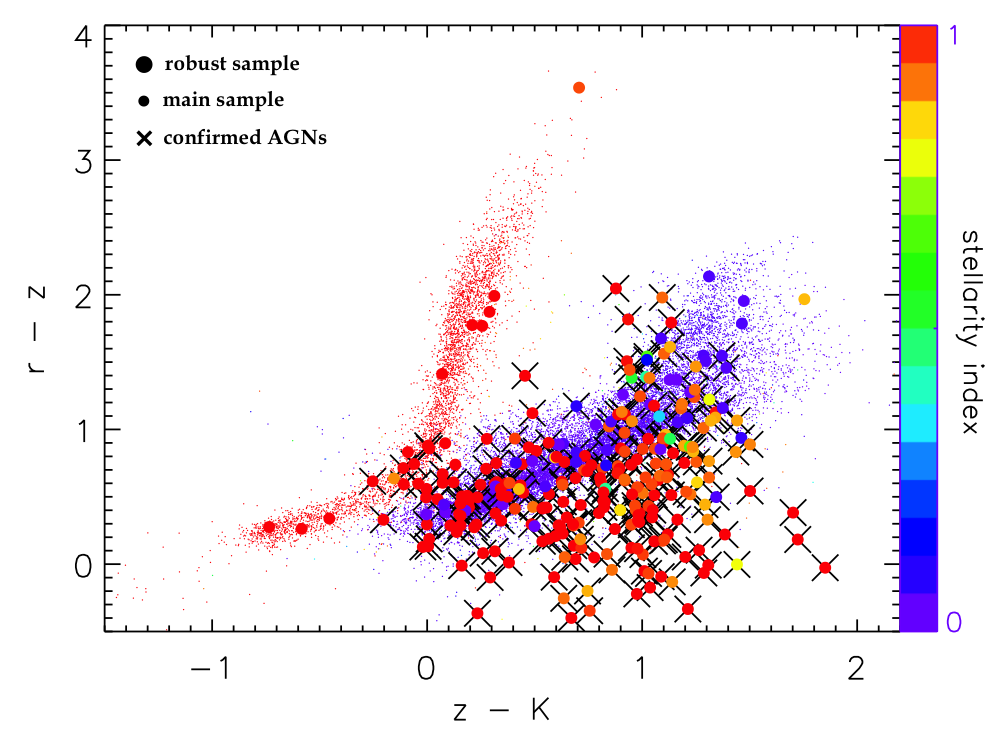}}
\caption{\emph{r-z} vs. \emph{z-K} diagram for the 296 out of 299 AGN candidates in the \emph{robust sample} (large dots) for which $r$, $z$, $K$ magnitudes and stellarity index are available. Small dots represent all the objects detected in the VST-COSMOS field for which \emph{HST} stellarity indices and optical/NIR color information are available, and are shown as a reference population. Both large and small dots are color-coded according to their stellarity index (right vertical axis). Crosses indicate X-ray emitting variable AGN candidates confirmed by their X-ray properties. It is apparent that the small dots in the plot define two distinct loci: one for stars (red), and one for galaxies (violet). A third class of objects, i.e., quasar-like AGNs, is identified from the large dots, corresponding to sources approximately near or below the galaxy locus but mostly with stellarity indices typical of compact sources. Approximately 6\% of the \emph{main sample} sources shown in Fig. \ref{fig:rzk} are located in the above-mentioned AGN locus, but are not part of our \emph{robust sample} of AGN candidates as they are below the variability threshold in the diagram shown in Fig. \ref{fig:stdev_vs_mag}. None of these sources has an X-ray counterpart in the \emph{X-ray sample}, and 96\% of them are extended sources based on their stellarity. $K_s$ magnitudes are by \citet{mcCracken}.}\label{fig:rzk}
\end{figure}

\citet{nakos} show that the use of the \emph{r-z} vs \emph{z-K} diagram is very effective in distinguishing stars from extended galaxies. The former form a tight sequence, while the latter tend to occupy a bluer NIR scattered region\footnote{Variable sources can change on different scales in different bands, but the effect is minimized by choosing bands close to each other \citep[e.g.,][]{simm}.}. Our \emph{r-z} vs. \emph{z-K} diagram is shown in Fig. \ref{fig:rzk}; color information is available for 296 out of 299 sources. When we overplot our variable AGN candidates on this diagram, there is a large overlap with the galaxy region that extends to redder colors, although they are on average more compact than galaxies, as revealed by their stellarity index. 

Based on this, and the stellar locus limit introduced by \citet{nakos}, we exclude from our \emph{robust sample} nine stellar sources (3\% of 299 variable AGN candidates) that lie on the stellar sequence far from the galaxy locus. A few additional AGN candidates lie in the overlap region between stars and galaxies, so we cannot assess their nature based on this diagram alone. We note that the number of stars detected over our three-year survey is significantly higher than in our first five-month analysis, where we detected no variables on the stellar locus of the \emph{r-z} vs. \emph{z-K} diagram, and is also larger than the fraction of stars found in the first 2 sq. deg. of the VST-CDFS survey, analyzed in \citet{falocco}. This is due to some extent to the longer temporal baseline and the fainter magnitude limit, but mostly to our more relaxed selection criterion (95th percentile threshold, see Section \ref{section:sample}) compared to the one used in our previous works, which allowed lower contamination.

Near-infrared photometry allows to separate stars from galaxies better than traditional optical colors (e.g., \emph{U-B} vs. \emph{B-V} diagrams), and to approximately identify an AGN locus, where point-like sources with colors typical of galaxies are located. Nonetheless, we do see a non-negligible fraction of variable AGNs --particularly optically redder and/or fainter ones, where the host galaxy contamination is more severe or where the nucleus is obscured-- whose colors and stellarity are consistent with those of inactive galaxies and hence are hard to identify by means of this diagram alone. A caveat here is that this diagnostic is prone to larger contamination than the diagnostics based on X-ray properties, since stars, galaxies, and AGNs have significant overlaps. As a consequence, we only use it to assess the extent of stellar contamination, which corresponds to the above mentioned 3\%.

\subsubsection{Mid-infrared diagnostic}
\label{section:lacy_donley}
\begin{figure}[tb]
 \center
   {\includegraphics[width=9cm]{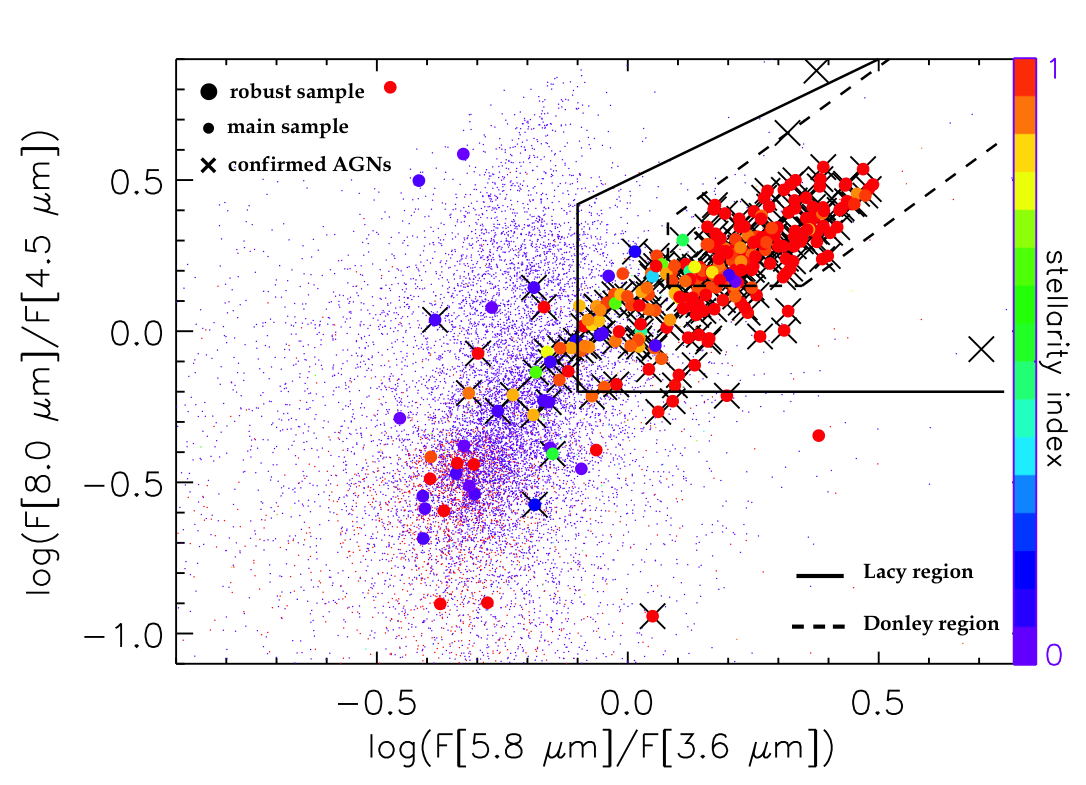}}
\caption{\footnotesize MIR diagram where colors are obtained as ratios of the fluxes in the four IRAC channels. Symbols are identical to Fig. \ref{fig:rzk}. The solid lines delineate the region where AGNs are typically found according to \citet{lacy07}, while the dashed lines define the less contaminated AGN region identified in \citet{donley}. A blob characterized by a high concentration of stars can be seen in the lower-left part of the diagram, while inactive galaxies tend to occupy the roughly vertical, scattered sequence defined by small violet dots.}\label{fig:irac}
\end{figure}

\citet{lacy04} proposed a color-color diagnostic based on the use of mid-infrared (MIR) fluxes (3.6, 4.5, 5.8, and 8.0 $\mu$m) from the Infrared Array Camera (IRAC; e.g., \citealt{fazio}) of the \emph{Spitzer Space Telescope} to select AGNs. The $8.0\mbox{ }\mu m/4.5\mbox{ }\mu m$ and the $5.8\mbox{ }\mu m/3.6\mbox{ }\mu m$ ratios allow separation of sources whose continuum emission is dominated by different components, such as stellar emission, dust reprocessing in star-forming regions, or nuclear dust heated by the central AGN. In Fig. \ref{fig:irac} several distinct loci are visible. Stars and low-redshift passive galaxies are characterized by bluer colors on both axes, and define the denser region that can be observed in the lower-left part of the plot. Star-forming galaxies with $z \lesssim 1.5$ are preferentially found on a roughly vertical sequence corresponding to colors with $5.8\mbox{ }\mu m/3.6\mbox{ }\mu m<0.1$ and $8.0\mbox{ }\mu m/4.5\mbox{ }\mu m>0.2$. Finally, quasars and AGN-dominated galaxies define a diagonal locus characterized by red colors on both axes. By defining color criteria, \citet{lacy04} delineated boundaries where AGNs generally are located on the diagram; this region was slightly modified in \citet{lacy07}, and we refer to it here and in the following (hereafter, the Lacy region). 

As shown by \citet{donley}, the Lacy region includes most galaxies with a MIR AGN contribution $\gtrsim 40\%$, but is not free from contamination by inactive galaxies. A more restrictive selection criterion was proposed by \citet{donley}, aiming at reducing heavy contamination by star-forming galaxies, which is particularly problematic at high redshift ($z \gtrsim 2$). \citet{donley} make use of simulations to construct IRAC colors of composite SEDs with different AGN contributions, in the redshift range 0--3, and show that, as the AGN contribution becomes dominant, the IRAC colors of the corresponding sources move onto a the locus where sources with perfect IRAC power-law SEDs would lie (hereafter, the Donley region).

Based on the analysis in \citet{falocco}, we show our version of the IRAC diagram in Fig. \ref{fig:irac}, which includes the AGN selection boundaries from both \citet{lacy07} and \citet{donley}, in order to allow a comparison. IRAC fluxes are from the COSMOS2015 catalog, and are available for 273 out of the 299 sources in our \emph{robust sample}. 
We find that $82\%$ (223/273) of the AGN candidates with IRAC fluxes lie within the Lacy region, including two sources confirmed \emph{ex novo}, i.e., not confirmed by any of the previous diagnostics. We caution that this does not constitute a purity estimate since the Lacy region is also known to include non-AGN contaminants. We assess the completeness of our variability selection with respect to MIR diagnostics, recovering 18\% (223/1235) and 55\% (138/249) of the sources that fall in the Lacy and Donley region, respectively. However, we note that the Lacy selection is affected by non-AGN contamination and it offers only a very loose lower limit (18\%) on the true completeness, while the Donley selection represent a more robust estimate (55\%).   

We find that variable sources mainly fall on the AGN power-law locus described above, supporting the view that a large fraction of the sample is comprised of AGN-dominated sources. However, a non-negligible fraction of variability-selected AGNs, confirmed by means of other diagnostics, fall outside the Lacy region. In these sources, the MIR host-galaxy emission likely dominates over the light contribution from the active nucleus (perhaps due to ongoing star formation), and hence AGNs are not easily distinguished from inactive galaxies by means of MIR colors. This confirms that optical variability, when coupled with multiwavelength photometry, can be a powerful tool to identify faint AGNs when other color selection methods alone fail.

\section{Discussion}
\label{section:discussion}
In this work we investigated the performance of optical variability as an AGN detection method, extending the monitoring baseline from the first five months already studied in \cite{decicco} up to three years, and the magnitude range 0.5 mag fainter. The larger baseline, coupled with the wealth of multiwavelength data available in the COSMOS field, allow significant improvements with respect to our previous investigations in COSMOS and the CDFS.
In particular, we are able to select a \emph{robust} sample of 299 AGN candidates, on the basis of their optical variability, and employ several multiwavelength diagnostics to confirm and characterize their nature. Table \ref{tab:confirmed} collects and summarizes the results obtained by means of the various diagnostics described in the previous sections, while Table \ref{tab:validation} contains a detailed list of all the AGN candidates in our \emph{robust sample}, with relevant information about them.

\begin{table}[tb]
\caption{Confirmed sources in the \emph{robust sample} of AGN candidates. We include the number of sources confirmed by individual diagnostics (lines 3 to 6), as well as the number of sources (when not null) confirmed only by a specific combination of diagnostics, or by a single one of them (lines 7 to 14).} 
\label{tab:confirmed}      
\centering
\begin{tabular}{c c}
\hline confirmed AGNs & 256 (86\% of 299)\\
confirmed stars & 9 (3\% of 299)\\
\hline
\ X/O validation (X) & 248 (83\% of 299)\\
spectroscopic validation (S) & 225 (75\% of 299)\\
Lacy region validation (L)& 223 (75\% of 299)\\
Donley region validation (D) & 138 (46\% of 299)\\ 
\hline
\ S+X+L+D validation & 134\\
S+X+L validation & 65\\
X+L+D validation & 3\\
S+X validation & 26\\
X+L validation & 14\\
L+D validation & 2\\
only X validation & 6\\
only L validation & 6\\
\hline classified sources & \\
with no X-ray counterpart & 8\\
\hline
\end{tabular}
\end{table}

\begin{figure*}[tb]
\centering
            {\includegraphics[width=\textwidth]{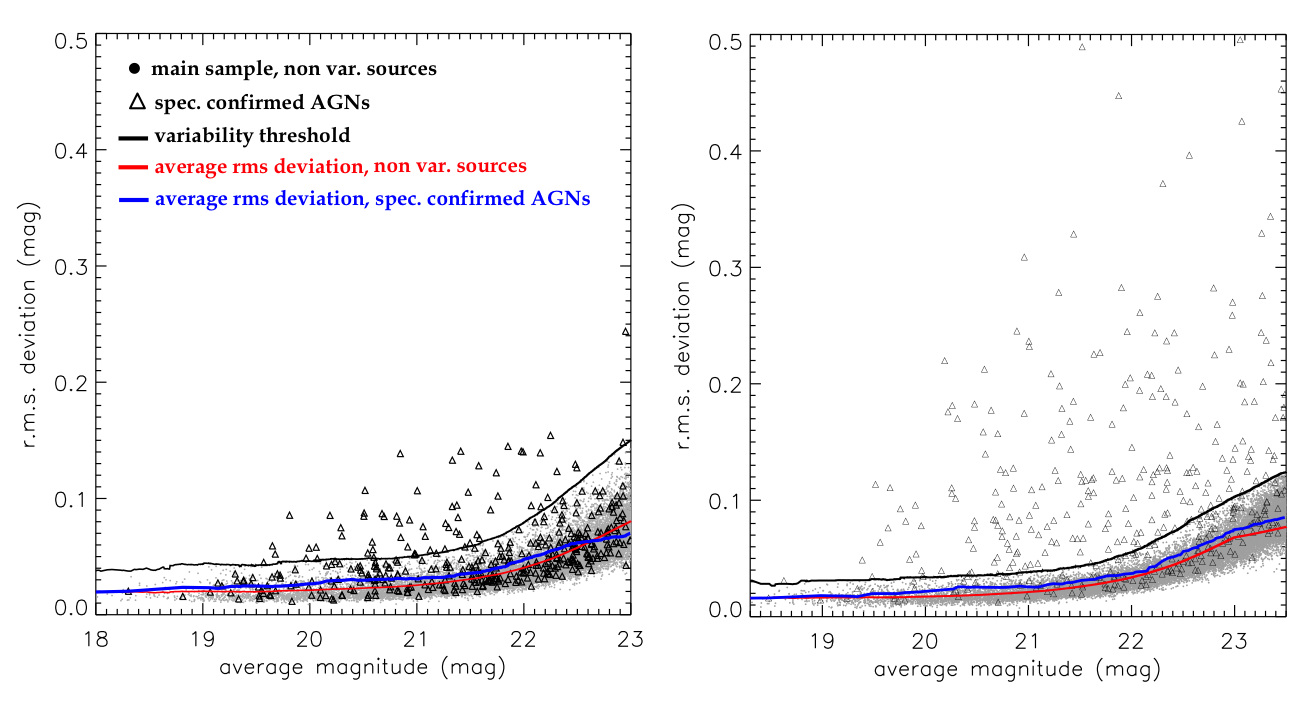}}
\caption{\footnotesize Light curve r.m.s. as a function of the average magnitude for all the non-variable sources in the \emph{main sample} (small dots) and for those with an X-ray counterpart and that are spectroscopically confirmed to be AGNs (triangles), from our five month analysis \citep[][\emph{left panel}]{decicco} and the present analysis (\emph{right panel}). In the left panel the vertical axis has been rescaled to the one in the right panel, while magnitudes are limited to \emph{r}(AB) $\leq$ 23.
The red and blue curves} represent the running average of the r.m.s. deviation for the two subsamples of sources, respectively. In the present work we find 59\% of the sources in the second subsample above the variability threshold (black line), while they were only 15\% in our former analysis.\label{fig:completeness}
\end{figure*}

Here follows a list of our main findings, together with a comparison with the results from our previous analysis described in \citet{decicco}.
\begin{enumerate}[i.]
\item We validate as AGNs 256 sources in the \emph{robust sample}, yielding a purity of $86\%$ (256/299), and demonstrating the effectiveness of optical variability as an AGN selection method. This represents an improvement compared to the $81\%$ value we obtained in \citet{decicco}, particularly considered the adoption of a more relaxed threshold (95\% instead of the $3\sigma$ used in our previous work) and 0.5 magnitude fainter detection threshold. In Table \ref{tab:validation} we report the percentile each source belongs to; purer samples can be obtained by choosing more restrictive thresholds (e.g., 99th percentile). \\
Regarding the contamination, our \emph{robust sample} includes two sources which are spectroscopically classified as stars in the \emph{Chandra} catalog; these two sources, together with seven additional X-ray undetected ones, lie on the stellar sequence in the diagram shown in Fig. \ref{fig:rzk}. Among the seven stellar candidates with sufficient MIR color information to plot in Fig. \ref{fig:irac}, six fall within the expected stellar locus. Importantly, none of these sources are  classified as AGNs by any of the diagnostics we used.
As a consequence, we classify all nine of them as stars, implying a stellar contamination of $3\%$ (9/299). This leaves 34 sources in the \emph{robust sample} with no classification. In the worst-case scenario where all of them are spurious, the contamination rate would be $14\%$. The 34 non-confirmed AGNs are mostly faint, only 10 of them having a magnitude \emph{r}(AB) $< 23$. 

In Section \ref{section:sample} we mentioned that the identification and analysis of SNe in VST-COSMOS data is currently ongoing (Ragosta et al., in prep.) but, at present, there are no identified SNe in our sample of AGN candidatates.

We note that three of the 34 sources do not have a counterpart in any of the known COSMOS catalogs, even when we search for them using a larger matching radius than 1\arcsec. These sources were already part of our sample of AGN candidates in \citet[][see Discussion there]{decicco}; they have average magnitudes $21.9 <$ \emph{r}(AB) $< 22.4$, which is well above \emph{HST} detection limit. Further investigation is necessary in order to unveil their nature.

We point out that 11 sources belonging to the \emph{robust sample} lie on the edges of the field, very close to the edge areas that we masked, or very close to other masked regions. Thus they are very likely to be spurious and we could have excluded them from our \emph{robust sample} if we had enforced a stricter masking policy. We had already noticed these sources and their positions at the time we obtained the \emph{robust sample}, but we chose not to exclude them \emph{a priori} because we did not want to introduce a bias in the sample.
Their exclusion would reduce the \emph{robust sample} to 288 AGN candidates, yielding a potential confirmation of $89\%$ and contamination ranging from $3\%$ to $11\%$, depending on the nature of the unclassified sources.
Three of the remaining 23 sources with no classification based on the used diagnostics are classified as stars in the COSMOS ACS catalog. 

If we consider that the analyzed sky area, not including masked regions, corresponds to $\approx 0.83$ sq. deg., the number of confirmed AGNs returns a density of $\approx 308$ AGNs per square degree.

\item We compute the completeness of AGNs in the \emph{robust sample} with respect to the most reliable sample of AGNs available in COSMOS, i.e., sources with an X-ray counterpart and a spectroscopic classification as AGN; we obtain $59\%$ completeness. This percentage is remarkably higher than the $18\%$\footnote{In \citet{decicco} we computed the completeness with respect to AGNs confirmed by spectroscopic and/or X-ray properties. Here we limit to spectroscopy and do not take into account X-ray properties. This explains why in our former work we report a completeness of 15\% instead of the 18\% we have just mentioned.} completeness obtained in \citet{decicco}. This means our result improved by a factor of 3.3. In Fig. \ref{fig:completeness} we show the improvement obtained in this work in the variability detection for X-ray AGNs compared to \citet{decicco}, by reporting two diagrams showing the completeness for both our past and present analyses. In the five-month analysis (left panel) VST sources with an X-ray counterpart exhibited, on average, a higher variability than the whole sample of VST sources, although $> 80\%$ of the sources with an X-ray counterpart fell below our variability threshold. At the time we predicted, based on the red-noise variability typical of AGNs, that a longer baseline would have returned a larger sample of sources above the variability threshold, and hence a much higher completeness (see Discussion and Fig. 8 in \citealt{decicco}); this is indeed what we found in the present analysis (right panel), where the completeness rises to 59\% and the fraction of sources below the threshold hence drops from $>80\%$ to 41\%.

The completeness for unobscured AGNs is now 82\%, while it is 18\% for obscured ones. In \citet{decicco} we report a 25\% and 6\% completeness, respectively. We note that our past analysis was limited to sources with \emph{r}(AB) $\le 23$ mag, while here we go half a magnitude deeper. Anyway, we verified that the results are not affected by the different magnitude threshold, and would be the same if we adopted the same cut. This suggests that the improvement in the detection rate is an effect of the longer baseline, and Fig. \ref{fig:completeness} confirms that.

The fraction of unobscured AGNs we retrieve is eight times larger than that of obscured ones. This is expected, since optical variability is biased towards the first class of AGNs, as we observe directly their inner regions \cite[e.g.,][and references therein]{padovani}.

Taking the cue from what we did in \citet{decicco}, we also computed the completeness in four magnitude bins, from \emph{r}(AB) $= 20$ to $23.5$ mag. We obtain 80\%, 66\%, and 53\%, respectively, for the first three bins, which are one magnitude-sized. Similarly to \citet{decicco}, these fractions are higher for brighter sources. The corresponding values\footnote{These fractions cannot properly be compared with the ones reported in \citet{decicco} due to the stricter definition here adopted for the completeness. Anyway, the large difference between the values reported for each bin is clearly dependent on the longer baseline here available, as already mentioned above.} in our previous work are 26\%, 23\%, and 5\%. The remaining bin is half a magnitude-sized, and has no correspondance in our previous analysis; the completeness, in this case, is 45\%.

In Section \ref{section:X} we mentioned that the completeness with respect to the AGNs confirmed by the X/O diagram is 37\%. We note that the \emph{X-ray sample} includes both unobscured and obscured AGNs, and that AGN selection based on optical variability is biased against the latter class. This is a possible explanation for the 37\% completeness that we obtain.

\item In Section \ref{section:sample} we stated that our \emph{robust sample} does not include 93 sources flagged 3 because of their doubtful variability. Based on the diagnostics we used throughout the present work, we would confirm as AGNs 12 of them. This number includes three sources confirmed only by the Lacy diagnostic which, as we mentioned in Section \ref{section:lacy_donley}, is affected by non-AGN contamination; as a consequence, the number of sources labeled 3 and with a reliable confirmation of their AGN nature is 9/93 (10\%), indicating that we are not biasing our results when removing sources affected by aesthetic or photometric problems.
\end{enumerate}

\begin{figure}[tb]
\centering
            {\includegraphics[width=8.9cm]{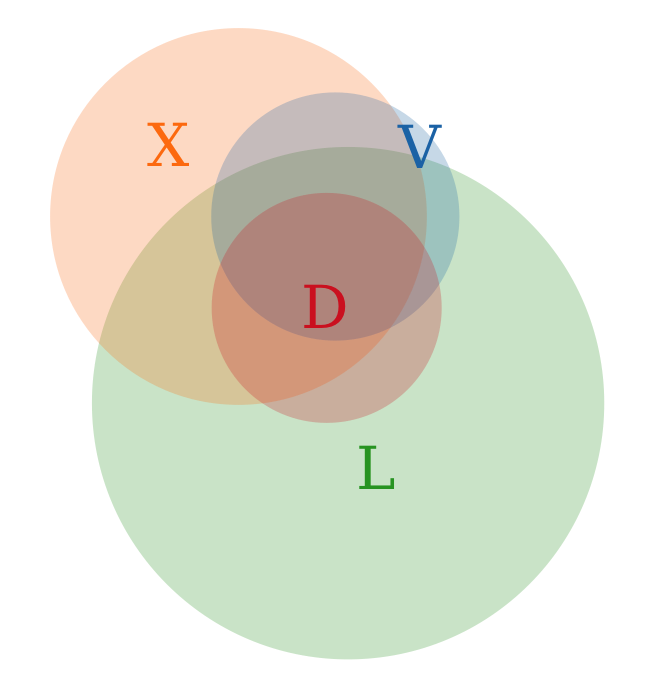}}
\caption{\footnotesize Venn diagram combining together the samples of AGNs selected by means of different diagnostics: optical variability (V), X-ray properties (X, from the X/O diagram), Lacy region (L), and Donley region (D).
The circle sizes, as well as the overlap sizes, are quantitatively correct. 
The sample of optically variable sources includes the 34 sources that are not confirmed by any other diagnostics, while it does not include the nine sources confirmed as stars (see main text). It is apparent that the variability-selected sample largely overlaps both with the X-ray selected and the Lacy-selected AGNs, while the overlap with the Donley-selected sample is only partial.}\label{fig:venn}
\end{figure}

In Fig. \ref{fig:venn} we show a Venn diagram summarizing the main results obtained by the various diagnostics used to validate our sources. The sample selected by means of optical variability is largely in overlap with both X-ray and IR ones, yet partly complementary to both. This is due to our relatively bright limit and the approach adopted in this work, which selects bright quasar-like AGNs; other works  show that addressing the problem of the host galaxy contamination through higher resolution instruments \citep[e.g.,][Pouliasis et al., in prep.]{Villforth1,sarajedini11,pouliasis} or through image subrtaction techniques \citep[e.g.,][]{botticella17} allows recovering of complementary samples of low-luminosity or X-ray faint AGNs.  In any case, variability allows to confirm AGN candidates identified by means of less robust/more contaminated diagnostics \citep[e.g.,][]{lacy07}.

\begin{figure}[bt]
 \center
   {\includegraphics[width=9cm]{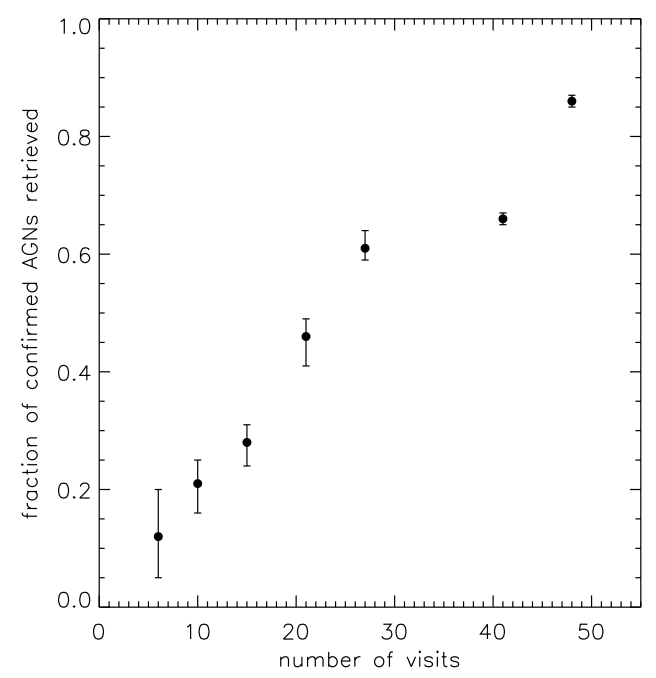}}
\caption{\footnotesize Fraction of confirmed AGNs retrieved selecting an increasing number of visits over a fixed baseline of 3.3 yr, which corresponds to the baseline covered in the present work, and is the longest sampled timescale for VST-COSMOS data to date. Each point is the average result of 10 random simulations, the error bars being the corresponding thresholds for the 10th and the 90th percentiles. The plot shows that, as the number of visits increases, we are able to retrieve a larger fraction of AGNs. A denser sampling leads to a reduction in the size of the error bars, as an effect of the reduced differences among the possible patterns that can be chosen for the simulation.}\label{fig:epoch_test}
\end{figure}

The 256 confirmed AGNs were found analyzing data from 54 visits over a 3 yr baseline. In order to assess how the sampling cadence could affect the AGN detection efficiency of future surveys, we made some tests, varying the number of visits over a fixed baseline of 3.3 yr. When possible, we included visits from each of the three seasons, in order to obtain a coverage as homogeneous as possible. Consistent with what we did in the present work (see Section \ref{section:selection}), we always required sources to be detected in at least half the visits. We show the results of this test in Fig. \ref{fig:epoch_test}. The test shows how significantly a denser sampling affects the detection efficiency; this effect is often underestimated when planning monitoring campaigns as it is uncorrectly assumed that just increasing the baseline is sufficient to increase the number of detections, even if this means dramatically reducing the sampled cadence.

Sections \ref{section:introduction} and \ref{section:selection} highlighted that the present work extends the analysis of \citet{decicco} to a longer baseline, and uses a different approach to correct magnitudes (reference visit vs. growth curves), select the sample of sources to include in the analysis (detection required in at least 50\% vs. 20\% of the visits), define a variability threshold (95th percentile vs. $3\sigma$  threshold). The present work also makes use of a $\sigma$-clipping algorithm when computing $\langle{\mbox{mag}}^{lc}\rangle$ and $\sigma^{lc}$ from the light curve of each source. It is thus worth comparing briefly the results obtained in our previous work with those we would obtain if we used this new approach in the analysis of the same dataset as in \citet{decicco}, consisting of 27 visits over the first season of the VST survey of COSMOS\footnote{We note that visit 27 is one of those excluded from the present work (see Section \ref{section:aperture} and Fig. \ref{fig:epoch_selection}). For the sake of consistency, we included this visit in the analysis performed to compare the results obtained with the two approaches over the same season. For the same reason, the sample of sources is now cut to \emph{r}(AB) mag $\le$ 23 mag, as in \citet{decicco}.}. 

In \citet{decicco} we obtained a sample of 83 AGN candidates (hereafter, \emph{sample 1}), and confirmed as AGNs 67 (81\%) of them. Following the new approach outlined in the present work, we would obtain a sample of 129 AGN candidates (hereafter, \emph{sample 2}), and confirm as AGNs 101 (78\%) of them, thus obtaining consistent results. In \emph{sample 2} we recover 84\% of the AGN candidates and 90\% of the confirmed AGNs in \emph{sample 1}. The constraint on the number of detections does not significantly affect the result, since most of the sources in each of the inspected samples were detected in almost all the visits. We expect \emph{sample 2} to be larger than \emph{sample 1} due to the different threshold adopted, which now roughly corresponds to a $1.5\sigma$ threshold, rather than $3\sigma$. The $\sigma$-clipping algorithm, on the other side, reduces the inclusion of spurious sources in the sample of AGN candidates.

There is one major factor to take into account, i.e., the red-noise type variability of AGNs. Indeed, the complete analysis over the 3.3 yr baseline returns 99\% of the confirmed AGNs from \emph{sample 1} and that do not fall in areas that in the new analysis we chose to mask. It also shows that there is only one AGN confirmed in \emph{sample 1} and not retrieved in the \emph{robust sample} because it is below the variability threshold. All this proves that the adopted approach, coupled with the longer baseline, leads to improved results with respect to \citet{decicco}, as also shown by the larger values obtained for purity and completeness.

In Section \ref{section:vst} we mentioned that so far our work has focused on $r$-band data; a complementary analysis of COSMOS $g$- and $i$-band variability, together with a multiband analysis combining $g$, $r$, and $i$ data, will be presented in a forthcoming paper. This will offer a chance to investigate correlated variability in different bands, together with the dependence of AGN variability on the specific wavelength range, and also average and time-dependent color selection \citep[e.g.,][]{richards15}; plus, the multiband analysis will retrieve more robust samples of candidates, simultaneously varying in multiple bands, thus minimizing contributions from contaminants.

\section*{Acknowledgements}
We acknowledge support from CONICYT grants Basal-CATA Basal AFB-170002 (D.D., F.E.B.), the Ministry of Economy, Development, and Tourism's Millennium Science Initiative through grant IC120009, awarded to The Millennium Institute of Astrophysics, MAS (D.D., F.E.B., G.P.), the ASI-INAF agreement n.2017-14-H.0.s (M.P., F.V., G.C.), the Italian Ministry of Foreign Affairs and International Cooperation (MAECI Grant Number ZA18GR02) and the South African Department of Science and Technology's National Research Foundation (DST-NRF Grant Number 113121) as part of the ISARP RADIOSKY2020 Joint Research Scheme (M.V.).

\bibliographystyle{aa}
\bibliography{aa2019_35659}

\clearpage
\onecolumn
\begin{landscape}
\begin{longtable}{c c c c c c c c l c r}
\caption{List of the 299 optically variable sources in the \emph{robust sample}. Column meanings: (1): identification number; (2) and (3): right ascension and declination (J2000); (4): average VST \emph{r}(AB) magnitude; (5): r.m.s. of the light curve; (6): \emph{SExtractor} stellarity index; (7): quality label; (8): percentile: the number indicates that the $\sigma^{lc}$ of the source is above the threshold of the corresponding percentile; (9): redshift: mostly spectroscopic redshifts from \citet{marchesi}; when not available, spectroscopic redshift from \citet{hasinger}, marked by a star; or spectroscopic redshift from \citet[][box]{lilly}; or  photometric redshift from \citet[][diamond]{marchesi}; or photometric redshift from \citet[][triangle]{Ilbert1}; (10): number of visits where the source is detected; (11): source classification. The classification index is the sum of different numbers corresponding to the following key:\\
1 = confirmed AGN through spectroscopy/SED; 2=confirmed AGN through X/O diagram; 4 = confirmed AGN through IRAC color diagram, Lacy region; 8 = confirmed AGN through IRAC color diagram, Donley region; 0 = non-classified; -1 = star.}\label{tab:validation}\\
\hline source ID & RA J2000 (deg) & Dec J2000 (deg) & avg $r(AB)$ mag (mag) & ltc r.m.s. (mag) & stellarity & quality label & percentile & redshift & \# of detections & classification\\
(1) & (2) & (3) & (4) & (5) & (6) & (7) & (8) & (9) & (10) & (11)\\
\endfirsthead
\caption{Continued.} \\
\hline source ID & RA J2000 (deg) & Dec J2000 (deg) & avg $r(AB)$ mag (mag) & ltc r.m.s. (mag) & stellarity & quality label & percentile & redshift & \# of detections & classification\\
(1) & (2) & (3) & (4) & (5) & (6) & (7) & (8) & (9) & (10) & (11)\\
\hline
\endhead
\hline
\endfoot
\hline
\  1     & 149.68732 &  1.7191443  & 20.26  &  0.11   & 0.90   & 1 &  99     & 1.35    &   54      & 15       \\    
\  2     & 150.07553  & 1.7199465  & 23.23   & 0.11   & 0.36   & 2 &  95      &0.868$\triangle$       &   46  &  0       \\
\  3     & 149.99166  & 1.7243215  & 20.68   & 0.11   & 0.93   & 1 &  98      & 1.625   &   52      & 15       \\
\  4     & 149.84359  & 1.7331874  & 23.36   & 0.12   & 0.97   & 2 &  95      & 0.059$\triangle$       &   47      &  0       \\
\  5     & 150.39678  & 1.7357309  & 22.47   & 0.11   & 0.97   & 1 &  98      & 2.766   &   54      & 15       \\
\  6     & 150.12141  & 1.7397895  & 23.49   & 0.13   & 0.01   & 1 &  96     & 0.723$\triangle$       &   36      &  0       \\
\  7     & 150.27996  & 1.7438591  & 23.08   & 0.20   & 0.43   & 2 &  99      & 0.763   &   51      &  3      \\
\  8     & 150.05048  & 1.7444366  & 22.20   & 0.21   & 0.72   & 1 &  99      & 1.149   &   52      &  7     \\
\  9     & 149.77079  & 1.7466641  & 21.53   & 0.14   & 0.92   & 2 &  99      & 1.327   &   54      &  2     \\
\  10    & 149.64483  & 1.7506337  & 19.52   & 0.11   & 0.87   & 1 &  99      & 1.9     &   54      & 15    \\
\  11    & 150.21676  & 1.7511995  & 20.03   & 0.04   & 0.03   & 2 &  96      & 0.1679$\Box$       &   54      &  0       \\
\  12    & 150.26242  & 1.7515061  & 23.27   & 0.28   & 0.85   & 1 &  99      & 1.676   &   44      & 15    \\
\  13    & 150.25924  & 1.7573333  & 22.71   & 0.17   & 0.02   & 2 &  99      & 0.965   &   54      &  6      \\
\  14    & 150.10740  & 1.7592058  & 22.73   & 0.11   & 0.98   & 1 &  98      & 3.949   &   51      &  2     \\
\  15    & 150.17663  & 1.7595157  & 22.20   & 0.12   & 0.83   & 1 &  99      & 1.161   &   54      & 15    \\
\  16    & 150.53564  & 1.7649016  & 21.81   & 0.17   & 0.94   & 1 &  99      & 2.204   &   54      & 15    \\
\  17    & 150.49173  & 1.7725595  & 21.86   & 0.06   & 0.16   & 1 &  97      & 0.833   &   54      &  7     \\
\  18    & 150.48517  & 1.7836715  & 22.92   & 0.11   & 0.98   & 1 &  97      & 2.727$\triangle$     &   53      & 14     \\
\  19    & 150.22784  & 1.7869356  & 23.07   & 0.11   & 0.31   & 1 &  95      & 1.681   &   52      &  7     \\
\  20    & 150.59603  & 1.7874690  & 21.52   & 0.12   & 0.89   & 1 &  99      & 1.246   &   54      & 15    \\
\  21    & 150.19511  & 1.7938135  & 22.98   & 0.27   & 0.51   & 1 &  99      & 1.867   &   54      & 15    \\
\  22    & 150.13174  & 1.7993850  & 21.10   & 0.07   & 0.93   & 1 &  98      & 1.679   &   52      & 15    \\
\  23    & 150.48525  & 1.8030591  & 22.56   & 0.12   & 0.85   & 1 &  99      & 0.957   &   54      & 15    \\
\  24    & 149.83469  & 1.8176868  & 22.80   & 0.22   & 0.98   & 1 &  99      & 0.735   &   50      & 15    \\
\  25    & 150.36590  & 1.8292915  & 21.82   & 0.38   & 0.91   & 1 &  99      & 0.958   &   54      &  6      \\
\  26    & 149.82193  & 1.8386822  & 21.22   & 0.21   & 0.93   & 1 &  99      & 1.351   &   52      & 15    \\
\  27    & 149.81686  & 1.8467454  & 21.03   & 0.08   & 0.95   & 1 &  98      & 1.035   &   53      &  7     \\
\  28    & 149.76952  & 1.8495966  & 22.98   & 0.26   & 0.90   & 1 &  99      & 2.242   &   52      &  7     \\
\  29    & 150.06795  & 1.8513342  & 21.29   & 0.12   & 0.89   & 2 &  99      & 1.134   &   52      &  7     \\
\  30    & 150.04684  & 1.8667762  & 22.00   & 0.15   & 0.97   & 1 &  99      & 2.412   &   52      &  7     \\
\  31    & 150.24309  & 1.8692006  & 20.89   & 0.06   & 0.91   & 1 &  97      & 2.021   &   53      & 15    \\
\  32    & 150.20827  & 1.8754041  & 20.64   & 0.18   & 0.91   & 1 &  99      & 1.147   &   54      & 15    \\
\  33    & 150.13916  & 1.8769595  & 20.81   & 0.06   & 0.90   & 1 &  97      & 0.831   &   54      & 15    \\
\  34    & 150.02547  & 1.8777086  & 23.10   & 0.18   & 0.93   & 1 &  99      & 1.796   &   51      &  7     \\
\  35    & 150.16181  & 1.8779243  & 22.41   & 0.24   & 0.98   & 1 &  99      & 1.444   &   54      & 15    \\
\  36    & 149.67421  & 1.8883298  & 20.43   & 0.04   & 0.89   & 1 &  95      & 1.784   &   54      & 15    \\
\  37    & 150.24520  & 1.9001845  & 20.19   & 0.22   & 0.91   & 1 &  99      & 1.56    &   50      & 15    \\
\  38    & 150.49135  & 1.9102721  & 22.34   & 0.11   & 0.97   & 1 &  98      & 2.234   &   51      & 15    \\
\  39    & 150.40095  & 1.9119384  & 21.47   & 0.07   & 0.95   & 1 &  98      & 2.279   &   51      & 15    \\
\  40    & 150.30891  & 1.9123230  & 22.43   & 0.31   & 0.88   & 1 &  99      & 1.483$\diamond$  &   52      &  6      \\
\  41    & 150.22938  & 1.9219723  & 23.48   & 0.13   & 0.01   & 2 &  96      & 0.543$\triangle$       &   29      &  0       \\
\  42    & 150.53943  & 1.9236345  & 20.63   & 0.05   & 0.03   & 1 &  97      & -       &   51      &  4      \\
\  43    & 150.58125  & 1.9269990  & 19.90   & 0.10   & 0.86   & 1 &  98      & 1.513   &   51      &  7     \\
\  44    & 150.32725  & 1.9286116  & 20.88   & 0.07   & 0.03   & 1 &  98      & 0.528   &   51      & 15    \\
\  45    & 150.11702  & 1.9298758  & 22.69   & 0.20   & 0.75   & 2 &  99      & 1.519   &   51      &  7     \\
\  46    & 149.98871  & 1.9324447  & 22.65   & 0.16   & 0.99   & 1 &  99      & 2.406   &   51      & 15    \\
\  47    & 150.43103  & 1.9352358  & 21.69   & 0.23   & 0.98   & 1 &  99      & 2.177   &   51      & 15    \\
\  48    & 150.35943  & 1.9364083  & 22.81   & 0.15   & 0.97   & 1 &  99      & 2.377   &   51      & 15    \\
\  49    & 150.22042  & 1.9539866  & 21.06   & 0.04   & 0.95   & 2 &  96      & 0.0     &   52      & -1     \\
\  50    & 149.67661  & 1.9563377  & 19.35   & 0.03   & 0.91   & 1 &  95      & 0.0$\triangle$       &   54      & -1     \\
\  51    & 150.27892  & 1.9595724  & 21.32   & 0.16   & 0.90   & 1 &  99      & 1.55    &   52      &  3       \\
\  52    & 150.35308  & 1.9607742  & 21.63   & 0.11   & 0.95   & 1 &  98      & 1.173   &   52      & 15    \\
\  53    & 150.25323  & 1.9661424  & 21.40   & 0.05   & 0.05   & 1 &  97      & 0.4263$\Box$       &   51      &  4      \\
\  54    & 150.38668  & 1.9665945  & 21.40   & 0.17   & 0.90   & 1 &  99      & 1.537   &   50      & 15    \\
\  55    & 150.44890  & 1.9767770  & 23.32   & 0.12   & 0.18   & 2 &  95      & 0.834$\triangle$       &   45      &  0       \\
\  56    & 150.57489  & 1.9767948  & 21.37   & 0.09   & 0.92   & 1 &  98      & 1.541   &   50      & 15    \\
\  57    & 150.31189  & 1.9779708  & 22.63   & 0.62   & 0.94   & 1 &  99      & 2.347   &   50      & 15    \\
\  58    & 149.92257  & 1.9792039  & 22.35   & 0.13   & 0.98   & 1 &  99      & 2.503   &   52      & 15    \\
\  59    & 149.63091  & 1.9794908  & 23.29   & 0.12   & 0.00   & 2 &  96      & 0.9989$\Box$       &   35      &  0       \\
\  60    & 149.67516  & 1.9827144  & 21.60   & 0.17   & 0.96   & 1 &  99      & 1.334   &   54      & 15    \\
\  61    & 149.77965  & 1.9843906  & 23.40   & 0.12   & 0.02   & 2 &  96      & 0.206$\triangle$       &   41      &  0       \\
\  62    & 150.46193  & 1.9846376  & 23.15   & 0.12   & 0.01   & 2 &  96      & 0.135$\triangle$       &   42      &  0       \\
\  63    & 150.21635  & 1.9887359  & 21.56   & 0.12   & 0.97   & 1 &  99      & 2.231   &   52      & 15    \\
\  64    & 150.15819  & 1.9911670  & 23.46   & 0.13   & 0.01   & 2 &  97      & 1.195$\triangle$       &   31      &  4      \\
\  65    & 150.25342  & 1.9966557  & 21.99   & 0.21   & 0.90   & 1 &  99      & 1.171   &   50      &  7     \\
\  66    & 149.95772  & 2.0030978  & 21.27   & 0.06   & 0.94   & 1 &  98      & 1.806   &   51      & 15    \\
\  67    & 150.07306  & 2.0035314  & 21.25   & 0.08   & 0.26   & 1 &  98      & 0.352   &   54      &  7     \\
\  68    & 150.19559  & 2.0044371  & 20.22   & 0.18   & 0.90   & 1 &  99      & 1.923   &   52      & 15    \\
\  69    & 150.50321  & 2.0047865  & 23.40   & 0.12   & 0.03   & 1 &  95      & 0.6606$\Box$       &   32      &  0       \\
\  70    & 149.99524  & 2.0066641  & 22.35   & 0.12   & 0.97   & 1 &  98      & 1.033   &   52      &  7     \\
\  71    & 150.46299  & 2.0090191  & 20.56   & 0.16   & 0.90   & 1 &  99      & 0.967   &   52      & 15    \\
\  72    & 150.42255  & 2.0141421  & 22.15   & 0.21   & 0.91   & 1 &  99      & 2.283   &   51      & 15    \\
\  73    & 150.28566  & 2.0146151  & 21.32   & 0.18   & 0.93   & 1 &  99      & 2.671   &   51      & 15    \\
\  74    & 150.05892  & 2.0151912  & 19.97   & 0.03   & 0.87   & 1 &  95      & 2.498   &   54      & 15    \\
\  75    & 149.81419  & 2.0163414  & 23.05   & 0.50   & 0.97   & 1 &  99      & 1.36    &   54      &  7     \\
\  76    & 149.71559  & 2.0165170  & 22.95   & 0.23   & 0.98   & 1 &  99      & 2.686   &   52      &  3       \\
\  77    & 149.95585  & 2.0280685  & 19.48   & 0.03   & 0.98   & 1 &  96      & 1.753   &   52      & 15    \\
\  78    & 149.88699  & 2.0336446  & 23.46   & 0.16   & 0.23   & 1 &  99      & -       &   48      &  0       \\
\  79    & 150.31195  & 2.0357509  & 21.58   & 0.08   & 0.63   & 1 &  98      & 0.971   &   51      &  3       \\
\  80    & 150.19448  & 2.0425058  & 20.13   & 0.05   & 0.88   & 1 &  97      & 0.0     &   53      & -1     \\
\  81    & 150.14566  & 2.0430791  & 20.91   & 0.08   & 0.91   & 1 &  98      & 1.178   &   53      & 15    \\
\  82    & 150.44603  & 2.0435056  & 21.13   & 0.11   & 0.91   & 1 &  98      & 1.171   &   51      &  7     \\
\  83    & 149.89512  & 2.0472080  & 22.21   & 0.12   & 0.98   & 1 &  99      & 2.198   &   51      & 15    \\
\  84    & 150.44369  & 2.0491068  & 20.42   & 0.07   & 0.89   & 1 &  97      & 0.669   &   51      & 15    \\
\  85    & 149.65632  & 2.0511320  & 23.48   & 0.18   & 0.89   & 1 &  99      & 1.854   &   48      &  7     \\
\  86    & 150.53574  & 2.0577803  & 22.09   & 0.07   & 0.97   & 1 &  97      & 1.195   &   51      &  3       \\
\  87    & 150.33439  & 2.0614565  & 20.26   & 0.11   & 0.88   & 1 &  98      & 0.904   &   52      & 15    \\
\  88    & 150.45768  & 2.0624709  & 21.58   & 0.10   & 0.98   & 1 &  98      & 2.325   &   51      & 15    \\
\  89    & 150.19474  & 2.0679277  & 20.48   & 0.08   & 0.03   & 1 &  98      & 0.552   &   53      & 15    \\
\  90    & 149.91055  & 2.0679895  & 23.28   & 0.20   & 0.97   & 1 &  99      & 1.091   &   50      & 15    \\
\  91    & 149.79440  & 2.0731069  & 23.33   & 0.14   & 0.98   & 1 &  99      & 2.668   &   52      & 15    \\
\  92    & 149.66920  & 2.0740675  & 20.15   & 0.04   & 0.03   & 1 &  96      & 0.34    &   54      &  7     \\
\  93    & 150.40639  & 2.0756581  & 23.05   & 0.20   & 0.89   & 1 &  99      & 1.257   &   49      &  7     \\
\  94    & 149.73378  & 2.0763561  & 22.98   & 0.14   & 0.96   & 2 &  99      & 2.042   &   49      &  7     \\
\  95    & 149.90957  & 2.0805839  & 20.74   & 0.10   & 0.90   & 1 &  98      & 2.802   &   52      & 15    \\
\  96    & 150.20798  & 2.0833580  & 19.76   & 0.09   & 0.87   & 1 &  98      & 1.236   &   53      & 15    \\
\  97    & 150.41830  & 2.0851684  & 20.82   & 0.04   & 0.03   & 2 &  96      & 0.424   &   51      &  7     \\
\  98    & 149.66359  & 2.0852282  & 20.47   & 0.06   & 0.89   & 1 &  97      & 1.22    &   54      &  7     \\
\  99    & 150.12289  & 2.0858446  & 23.50   & 0.14   & 0.03   & 2 &  98      & 0.6606$\Box$       &   45      &  0       \\
\  100   & 150.32426  & 2.0890822  & 23.31   & 0.24   & 0.99   & 1 &  99      & 1.797   &   50      &  3       \\
\  101   & 149.82267  & 2.0896669  & 22.45   & 0.21   & 0.91   & 1 &  99      & 2.051   &   54      &  7     \\
\  102   & 150.12558  & 2.0907611  & 23.22   & 0.12   & 0.98   & 2 &  96      & 0.0$\triangle$       &   52      &  0       \\
\  103   & 149.89795  & 2.0939249  & 22.28   & 0.20   & 0.98   & 1 &  99      & 1.921   &   51      & 15    \\
\  104   & 150.10203  & 2.1054810  & 19.85   & 0.05   & 0.86   & 2 &  98      & 2.288   &   54      & 15    \\
\  105   & 150.22434  & 2.1056295  & 21.51   & 0.10   & 0.04   & 2 &  98      & 0.74    &   52      &  3      \\
\  106   & 150.55902  & 2.1057491  & 22.83   & 0.17   & 0.97   & 1 &  99      & 2.323   &   50      & 15    \\
\  107   & 150.59573  & 2.1111357  & 22.02    & 0.10  & 0.94   & 2 &  98      & 0.899   &   51      & 15    \\
\  108   & 150.37361  & 2.1120309  & 20.85   & 0.13   & 0.95   & 1 &  99      & 1.914   &   51      & 15    \\
\  109   & 150.42092  & 2.1124419  & 23.27   & 0.19   & 0.78   & 1 &  99      & 1.263   &   49      &  6      \\
\  110   & 150.29241  & 2.1137305  & 22.95   & 0.11   & 0.87   & 1 &  97      & 1.828   &   53      &  7     \\
\  111   & 150.26170  & 2.1166520  & 21.43   & 0.07   & 0.98   & 1 &  98      & 2.572   &   52      & 15    \\
\  112   & 149.93092  & 2.1187187  & 19.66   & 0.11   & 0.88   & 1 &  99      & 2.194   &   54      & 15    \\
\  113   & 150.32828  & 2.1249531  & 22.25   & 0.12   & 0.95   & 1 &  99      & 1.774   &   52      &  7     \\
\  114   & 150.10559  & 2.1263062  & 21.96   & 0.24   & 0.95   & 1 &  99      & 0.961   &   54      & 15    \\
\  115   & 149.85448  & 2.1324290  & 23.03   & 0.11   & 0.30   & 1 &  97      & 0.955   &   50      &  2     \\
\  116   & 150.19901  & 2.1325192  & 19.81   & 0.08   & 0.89   & 1 &  98      & 2.159   &   54      & 15    \\
\  117   & 150.15847  & 2.1395817  & 20.70   & 0.08   & 0.92   & 1 &  98      & 1.826   &   54      & 15    \\
\  118   & 150.43540  & 2.1428146  & 22.88   & 0.14   & 0.03   & 1 &  99      & 0.968   &   51      &  7     \\
\  119   & 150.45186  & 2.1448420  & 20.89   & 0.25   & 0.91   & 1 &  99      & 1.298   &   51      & 15    \\
\  120   & 149.71124  & 2.1451715  & 20.73   & 0.13   & 0.90   & 1 &  99      & 1.53    &   54      & 15    \\
\  121   & 150.34592  & 2.1475287  & 20.91   & 0.04   & 0.91   & 1 &  95      & 1.256   &   54      &  7     \\
\  122   & 150.29726  & 2.1488036  & 20.97   & 0.04   & 0.92   & 1 &  96      & 3.328   &   53      & 15    \\
\  123   & 150.10874  & 2.1569871  & 23.40   & 0.17   & 0.96   & 2 &  99      & 1.814   &   44      &  7     \\
\  124   & 150.30261  & 2.1610864  & 20.70   & 0.05   & 0.90   & 1 &  97      & 1.817   &   53      & 15    \\
\  125   & 150.48442  & 2.1620498  & 22.34   & 0.19   & 0.98   & 1 &  99      & 1.639   &   53      &  7     \\
\  126   & 149.82791  & 2.1643113  & 20.48   & 0.08   & 0.89   & 1 &  98      & 1.156   &   52      & 15    \\
\  127   & 150.47551  & 2.1710079  & 23.46   & 0.12   & 0.00   & 2 &  95      & 1.269*       &   33      &  0       \\
\  128   & 149.89485  & 2.1744358  & 22.08   & 0.26   & 0.97   & 1 &  99      & 1.312   &   54      &  7     \\
\  129   & 150.42224  & 2.1754161  & 21.44   & 0.33   & 0.88   & 2 &  99      & 0.979   &   54      & 15    \\
\  130   & 150.57633  & 2.1813797  & 20.88   & 0.05   & 0.03   & 1 &  97      & 0.554   &   53      &  3      \\ 
\  131   & 150.53190  & 2.1888941  & 21.78   & 0.09   & 0.97   & 1 &  98      & 0.827   &   52      & 15    \\
\  132   & 150.19984  & 2.1908592  & 21.22   & 0.06   & 0.92   & 1 &  98      & 1.515   &   54      & 15    \\
\  133   & 150.37825  & 2.1964202  & 23.26   & 0.33   & 0.63   & 1 &  99      & 1.512   &   50      & 7     \\
\  134   & 149.91246  & 2.2003194  & 20.75   & 0.07   & 0.91   & 1 &  98      & 0.686   &   54      & 15    \\
\  135   & 150.04599  & 2.2011612  & 22.32   & 0.09   & 0.91   & 2 &  98      & 2.704   &   54      &  7     \\
\  136   & 150.21473  & 2.2042706  & 21.22   & 0.15   & 0.93   & 1 &  99      & 1.841   &   54      & 15    \\
\  137   & 150.13410  & 2.2045658  & 23.50   & 0.13   & 0.17   & 1 &  96      & 0.67$\triangle$       &   32      &  0       \\
\  138   & 150.08603  & 2.2097157  & 21.78   & 0.10   & 0.07   & 2 &  98      & 0.2386$\Box$       &   54      &  0       \\
\  139   & 150.02856  & 2.2099369  & 21.44   & 0.18   & 0.92   & 1 &  99      & 1.258   &   53      & 15    \\
\  140   & 150.42124  & 2.2166694  & 22.06   & 0.06   & 0.90   & 1 &  96      & 0.621   &   54     &  7     \\
\  141   & 149.73895  & 2.2206892  & 20.68   & 0.08   & 0.91   & 1 &  98      & 1.024   &   49      & 15    \\
\  142   & 150.58111  & 2.2210492  & 22.31   & 0.08   & 0.98   & 1 &  97      & 2.022   &   52      & 15    \\
\  143   & 149.87918  & 2.2258118  & 22.74   & 0.11   & 0.92   & 2 &  98      & 3.65    &   51      &  3      \\
\  144   & 149.82359  & 2.2282771  & 22.62   & 0.11   & 0.98   & 1 &  98      & 2.97    &   49      &  7     \\
\  145   & 150.27216  & 2.2300959  & 20.96   & 0.17   & 0.92   & 1 &  99      & 2.611   &   54      & 15    \\
\  146   & 150.00447  & 2.2370960  & 22.20   & 0.10   & 0.56   & 1 &  98      & 1.406   &   51      & 15    \\
\  147   & 150.09361  & 2.2372209  & 23.01   & 0.14   & 0.09   & 1 &  99      & 1.4373$\Box$       &   51      &  4      \\
\  148   & 149.89539  & 2.2394848  & 21.09   & 0.05   & 0.93   & 1 &  97      & 1.729   &   52      & 15    \\
\  149   & 150.34994  & 2.2461224  & 21.29   & 0.28   & 0.90   & 1 &  99      & 0.894   &   54      &  7     \\
\  150   & 150.44961  & 2.2464424  & 20.78   & 0.12   & 0.91   & 1 &  99      & 0.882   &   54      & 15    \\
\  151   & 149.74392  & 2.2497706  & 18.63   & 0.03   & 0.03   & 2 &  97      & 0.133   &   48      &  3      \\
\  152   & 150.47956  & 2.2531361  & 22.45   & 0.08   & 0.95   & 1 &  96      & 0.866   &   54      &  6      \\
\  153   & 149.82193  & 2.2546519  & 22.27   & 0.13   & 0.03   & 2 &  99      & 0.934   &   47      &  3      \\
\  154   & 150.01053  & 2.2559001  & 23.30   & 0.12   & 0.97   & 1 &  96      & 1.661   &   49      &  7     \\
\  155   & 149.77136  & 2.2583635  & 21.91   & 0.12   & 0.93   & 2 &  99      & 2.213   &   47      & 15    \\
\  156   & 149.99364  & 2.2585206  & 21.51   & 0.12   & 0.04   & 2 &  99      & 0.658   &   51      &  7     \\
\  157   & 150.01822  & 2.2594497  & 23.21   & 0.12   & 0.98   & 1 &  96      & 2.673   &   50      & 15    \\
\  158   & 150.42609  & 2.2600481  & 23.16   & 0.12   & 0.00   & 1 &  98      & 0.814*       &   46      &  0       \\
\  159   & 150.49208  & 2.2634375  & 23.50   & 0.16   & 0.96   & 2 &  99      & 0.835$\triangle$       &   48      &  0       \\
\  160   & 150.39264  & 2.2701100  & 23.38   & 0.14   & 0.96   & 1 &  98      & 2.158   &   49      &  7     \\
\  161   & 150.53616  & 2.2732704  & 21.77   & 0.08   & 0.57   & 1 &  98      & 1.081   &   54      &  7     \\
\  162   & 150.32195  & 2.2744886  & 22.31   & 0.13   & 0.88   & 2 &  99      & 0.3517$\Box$       &   54      &  4      \\
\  163   & 150.00927  & 2.2755091  & 20.58   & 0.14   & 0.85   & 2 &  99      & 0.85    &   52      & 15    \\
\  164   & 149.85155  & 2.2764230  & 23.37   & 0.13   & 0.98   & 2 &  98      & 3.372   &   44      &  3      \\
\  165   & 149.70633  & 2.2779600  & 21.61   & 0.12   & 0.19   & 1 &  99      & 0.731   &   54      &  7     \\
\  166   & 149.66680  & 2.2864436  & 21.88   & 0.45   & 0.92   & 1 &  99      & 1.028   &   54      &  7     \\
\  167   & 150.58184  & 2.2877456  & 21.81   & 0.12   & 0.96   & 1 &  99      & 1.343   &   54      & 15    \\
\  168   & 149.63869  & 2.2890013  & 20.72   & 0.09   & 0.94   & 1 &  98      & 0.231$\triangle$       &   54      & 12    \\
\  169   & 150.23628  & 2.2891123  & 21.58   & 0.11   & 0.96   & 1 &  98      & 2.076   &   54      & 15    \\
\  170   & 150.08439  & 2.2904945  & 22.80   & 0.28   & 0.87   & 1 &  99      & 2.101   &   51      &  7     \\
\  171   & 150.13800  & 2.2916745  & 19.80   & 0.04   & 0.03   & 2 &  96      & 0.185   &   52      &  2     \\
\  172   & 149.71529  & 2.2994018  & 23.49   & 0.19   & 0.97   & 1 &  99      & 1.191   &   46      &  7     \\
\  173   & 149.99392  & 2.3014085  & 20.29   & 0.10   & 0.88   & 1 &  98      & 1.796   &   51      & 15    \\
\  174   & 150.36670  & 2.3053423  & 21.94   & 0.20   & 0.98   & 1 &  99      & 1.187   &   54      &  7     \\
\  175   & 149.86340  & 2.3060702  & 23.45   & 0.13   & 0.98   & 2 &  97      & 0.74$\triangle$       &   44      & -1    \\
\  176   & 149.91975  & 2.3274421  & 20.82   & 0.09   & 0.91   & 1 &  98      & 1.454   &   51      & 15    \\
\  177   & 150.06456  & 2.3290367  & 22.42   & 0.18   & 0.98   & 1 &  99      & 2.45    &   52      & 15    \\
\  178   & 149.86802  & 2.3306993  & 20.70   & 0.15   & 0.90   & 1 &  99      & 1.478   &   49      &  7     \\
\  179   & 149.76351  & 2.3341496  & 21.41   & 0.14   & 0.93   & 1 &  99      & 1.137   &   47      & 15    \\
\  180   & 149.79165  & 2.3382080  & 21.93   & 0.07   & 0.87   & 1 &  97      & 1.008   &   49      &  3      \\
\  181   & 149.95514  & 2.3402663  & 19.65   & 0.03   & 0.87   & 1 &  96      & 0.01$\triangle$       &   52      & -1     \\
\  182   & 150.35361  & 2.3421398  & 20.74   & 0.07   & 0.94   & 1 &  98      & 1.708   &   54      & 15    \\
\  183   & 149.84949  & 2.3431028  & 23.10   & 0.25   & 0.98   & 2 &  99      & 0.792   &   48      & 14     \\
\  184   & 149.91949  & 2.3453929  & 21.85   & 0.12   & 0.91   & 1 &  99      & 3.015   &   51      & 15    \\
\  185   & 149.81295  & 2.3454732  & 22.26   & 0.07   & 0.93   & 2 &  96      & 1.812   &   48      &  7     \\
\  186   & 149.88329  & 2.3467476  & 21.94   & 0.09   & 0.92   & 1 &  98      & 1.021   &   51      &  7     \\
\  187   & 149.65292  & 2.3469199  & 23.07   & 0.43   & 0.97   & 1 &  99      & 1.188   &   54      &  7     \\
\  188   & 149.67864  & 2.3489133  & 23.47   & 0.17   & 0.91   & 1 &  99      & 2.017   &   45      &  7     \\
\  189   & 149.72076  & 2.3489995  & 22.73   & 0.20   & 0.94   & 1 &  99      & 1.53    &   52      &  6      \\
\  190   & 149.86798  & 2.3518587  & 20.45   & 0.04   & 0.03   & 2 &  96      & 0.346   &   51      &  7     \\
\  191   & 150.25560  & 2.3534942  & 23.03   & 0.25   & 0.94   & 1 &  99      & 2.975$\diamond$  &   52      &  6      \\
\  192   & 150.02012  & 2.3536062  & 22.06   & 0.08   & 0.98   & 1 &  98      & 2.663$\diamond$  &   52      & 14     \\
\  193   & 149.95061  & 2.3545852  & 23.16   & 0.36   & 0.15   & 1 &  99      & 0.919$\triangle$       &   49      &  0       \\
\  194   & 150.12366  & 2.3582506  & 21.01   & 0.09   & 0.05   & 1 &  98      & 0.728   &   52      &  7     \\
\  195   & 150.23181  & 2.3639768  & 21.18   & 0.08   & 0.94   & 1 &  98      & 1.936   &   54      & 15    \\
\  196   & 150.40667  & 2.3654220  & 22.25   & 0.28   & 0.98   & 1 &  99      & 2.032   &   54      & 15    \\
\  197   & 149.73882  & 2.3663495  & 23.00   & 0.14   & 0.93   & 2 &  99      & 1.65    &   47      &  3      \\
\  198   & 149.94552  & 2.3692673  & 21.52   & 0.49   & 0.94   & 1 &  99      & 0.909   &   52      & 15    \\
\  199   & 150.44442  & 2.3697490  & 22.20   & 0.19   & 0.44   & 1 &  99      & 0.891   &   54      &  7     \\
\  200   & 149.84814  & 2.3742413  & 21.01   & 0.23   & 0.96   & 1 &  99      & 2.735   &   48      & 15    \\
\  201   & 149.83285  & 2.3846825  & 22.34   & 0.24   & 0.98   & 1 &  99      & 2.247   &   47      & 15    \\
\  202   & 150.52679  & 2.3846957  & 22.97   & 0.12   & 0.86   & 1 &  98      & 1.299   &   52      &  3      \\
\  203   & 149.91690  & 2.3851880  & 19.96   & 0.08   & 0.88   & 1 &  98      & 1.131   &   52      & 15    \\
\  204   & 150.42194  & 2.3854774  & 23.26   & 0.24   & 0.98   & 1 &  99      & 1.514   &   46      &  3      \\
\  205   & 150.45038  & 2.3880675  & 23.42   & 0.20   & 0.97   & 1 &  99      & 1.075$\diamond$ &   45      &  6      \\
\  206   & 150.21076  & 2.3914776  & 22.85   & 0.10   & 0.60   & 1 &  96      & 3.095   &   51      & 15    \\
\  207   & 150.45848  & 2.3973237  & 22.20   & 0.08   & 0.86   & 2 &  98      & 2.89    &   54      & 15    \\
\  208   & 150.09155  & 2.3990645  & 22.37   & 0.14   & 0.99   & 1 &  99      & 2.47    &   52      & 15    \\
\  209   & 150.17396  & 2.4030030  & 23.16   & 0.14   & 0.43   & 1 &  99      & 0.979   &   46      & 15    \\
\  210   & 150.51163  & 2.4096278  & 21.03   & 0.11   & 0.92   & 1 &  98      & 0.986   &   54      & 15    \\
\  211   & 150.47206  & 2.4102167  & 20.85   & 0.11   & 0.79   & 1 &  98      & 0.668   &   54      & 15    \\
\  212   & 149.85622  & 2.4104209  & 20.78   & 0.04   & 0.94   & 2 &  95      & 0.0$\triangle$       &   46      & -1     \\
\  213   & 149.66026  & 2.4109485  & 22.60   & 0.13   & 0.10   & 1 &  99      & 1.161   &   53      & 15    \\
\  214   & 150.49569  & 2.4125739  & 21.75   & 0.08   & 0.87   & 1 &  98      & 1.369   &   54      & 15    \\
\  215   & 150.43815  & 2.4157843  & 20.80   & 0.06   & 0.91   & 1 &  97      &  2.02   &   54      & 15    \\
\  216   & 149.87067  & 2.4172606  & 22.54   & 0.12   & 0.98   & 2 &  99      & 1.528   &   50      & 15    \\
\  217   & 149.76069  & 2.4198974  & 22.08   & 0.12   & 0.94   & 1 &  99      & 1.101   &   50      &  3      \\
\  218   & 150.42437  & 2.4220652  & 18.51   & 0.03   & 0.03   & 1 &  95      & -       &   54      &  0       \\
\  219   & 150.38663  & 2.4223974  & 21.05   & 0.04   & 0.97   & 1 &  95      & -       &   54      &  0       \\
\  220   & 150.54404  & 2.4230778  & 23.19   & 0.18   & 0.01   & 1 &  99      & 0.957   &   46      &  3      \\
\  221   & 150.51395  & 2.4233098  & 20.08   & 0.03   & 0.96   & 1 &  95      & -       &   54      & -1     \\
\  222   & 150.12409  & 2.4313257  & 22.98   & 0.11   & 0.00   & 2 &  96      & 0.757*       &   32      &  0       \\ 
\  223   & 150.20898  & 2.4384817  & 21.60   & 0.05   & 0.96   & 1 &  95      & 3.715   &   52      & 15    \\
\  224   & 149.93588  & 2.4405710   & 22.34   & 0.13  & 0.87   & 1 &  99      & 0.993   &   54      & 15    \\ 
\  225   & 149.77695  & 2.4442782  & 23.12   & 0.12   & 0.97   & 2 &  98      & 4.171   &   50      &  3      \\
\  226   & 150.49916  & 2.4449390   & 19.56   & 0.06  & 0.88   & 1 &  98      & 2.03    &   54      & 15    \\
\  227   & 150.06217  & 2.4550396  & 21.53   & 0.07   & 0.14   & 1 &  98      & 0.73    &   52      &  7     \\
\  228   & 149.72048  & 2.4587871  & 23.47   & 0.16   & 0.59   & 2 &  99      & -       &   45      &  0       \\
\  229   & 150.32742  & 2.4608443  & 20.66   & 0.08   & 0.62   & 1 &  98      & 1.042   &   54      & 15    \\
\  230   & 150.03588  & 2.4645999  & 23.29   & 0.12   & 0.98   & 2 &  95      & -       &   50      &  0       \\
\  231   & 149.75093  & 2.4699039  & 21.16   & 0.06   & 0.51   & 2 &  97      & 0.658   &   51      &  7     \\
\  232   & 150.00452  & 2.4765675  & 20.20   & 0.03   & 0.98   & 1 &  95      & 0.0$\triangle$       &   54      & -1     \\
\  233   & 150.10669  & 2.4772543  & 22.03   & 0.06   & 0.98   & 1 &  96      & 0.0$\triangle$       &   52      & -1     \\
\  234   & 150.05872  & 2.4774000  & 20.48   & 0.18   & 0.91   & 1 &  99      & 1.258   &   54      & 15    \\
\  235   & 150.42870  & 2.4782022  & 23.21   & 0.11   & 0.02   & 2 &  95      & 0.168$\triangle$       &   45      & 12    \\
\  236   & 150.40702  & 2.4788544  & 21.92   & 0.18   & 0.32   & 1 &  99      & 0.367   &   53      & 15    \\
\  237   & 150.20884  & 2.4819287  & 19.96   & 0.04   & 0.87   & 1 &  97      & 3.333   &   52      & 15    \\
\  238   & 149.97694  & 2.4866390   & 23.08   & 0.15  & 0.93   & 1 &  99      & 2.089   &   51      & 15    \\
\  239   & 149.99068  & 2.4990856  & 23.39   & 0.12   & 0.80   & 2 &  95      & 2.022$\triangle$       &   32      &  0       \\
\  240   & 150.37096  & 2.5001018  & 23.44   & 0.13   & 0.98   & 2 &  96      & 0.0$\triangle$       &   39      &  0       \\
\  241   & 149.95562  & 2.5020388  & 22.54   & 0.17   & 0.96   & 1 &  99      & 1.458   &   53      &  7     \\
\  242   & 150.39866  & 2.5028107  & 23.41    & 0.13  & 0.02   & 2 &  96      & -       &   32      &  0       \\
\  243   & 150.56735  & 2.5035958  & 21.50   & 0.06   & 0.92   & 1 &  97      & 1.146   &   54      & 15    \\
\  244   & 149.93839  & 2.5059488  & 22.58   & 0.08   & 0.00   & 2 &  95      & 0.893   &   53      &  3      \\
\  245   & 150.54475  & 2.5072799  & 19.39   & 0.05   & 0.90   & 1 &  98      & 1.16    &   54      & 15    \\
\  246   & 149.92025  & 2.5142271  & 22.68   & 0.09   & 0.96   & 1 &  96      & 0.698   &   53      &  3      \\
\  247   & 150.45062  & 2.5173912  & 21.93   & 0.13   & 0.96   & 1 &  99      & -       &   51      &  0       \\
\  248   & 150.60068  & 2.5188092  & 22.63   & 0.12   & 0.42   & 1 &  99      & 0.669   &   53      &  3      \\
\  249   & 150.51820  & 2.5217408  & 22.92   & 0.13   & 0.77   & 1 &  99      &  2.779  &   53      & 15    \\
\  250   & 150.41591  & 2.5257835  & 22.57   & 0.10   & 0.98   & 1 &  98      &  1.444  &   53      &  3      \\
\  251   & 150.45356  & 2.5279411  & 21.26   & 0.13   & 0.91   & 1 &  99      &  1.935  &   54      & 15    \\
\  252   & 150.56357  & 2.5361171  & 22.46   & 0.10   & 0.02   & 2 &  98      &  0.496$\triangle$      &   53      &  0       \\
\  253   & 150.53936  & 2.5420789  & 22.50   & 0.10   & 0.97   & 1 &  98      &  1.716  &   52      &  7     \\
\  254   & 150.38608  & 2.5426687  & 22.41   & 0.14   & 0.87   & 2 &  99      &  -      &   52      &  0       \\
\  255   & 149.86034  & 2.5447442  & 23.35   & 0.22   & 0.98   & 1 &  99      &  1.596  &   51      & 15    \\
\  256   & 149.69586  & 2.5490825  & 22.77   & 0.10   & 0.14   & 2 &  96      &  1.776  &   41      &  7     \\
\  257   & 149.94219  & 2.5522457  & 23.44   & 0.12   & 0.00   & 2 &  95      &  0.6683$\Box$      &   28      &  0       \\
\  258   & 149.91067  & 2.5546511  & 20.85   & 0.05   & 0.65   & 1 &  97      &  0.753  &   54      &  7     \\
\  259   & 149.78568  & 2.5547887  & 23.13   & 0.14   & 0.93   & 1 &  99      &  1.798  &   52      &  7     \\
\  260   & 149.81148  & 2.5579622  & 23.35   & 0.34   & 0.72   & 1 &  99      &  2.152  &   47      &  7     \\
\  261   & 149.79633  & 2.5594275  & 22.72   & 0.11   & 0.96   & 1 &  98      &  1.541  &   52      & 15    \\
\  262   & 150.38282  & 2.5597507  & 23.45   & 0.45   & 0.48   & 1 &  99      &  2.066  &   33      & 15    \\
\  263   & 150.33449  & 2.5614370  & 20.26   & 0.18   & 0.88   & 1 &  99      &  1.836  &   54      & 15    \\
\  264   & 149.79622  & 2.5641002  & 21.02   & 0.04   & 0.91   & 1 &  96      &  0.704  &   53      & 15    \\
\  265   & 150.55593  & 2.5643882  & 22.35   & 0.09   & 0.97   & 1 &  98      &  1.143  &   54      & 15    \\
\  266   & 149.77797  & 2.5672597  & 22.09   & 0.18   & 0.79   & 2 &  99      &  0.72   &   54      &  6      \\
\  267   & 149.70362  & 2.5781197  & 21.00   & 0.24   & 0.89   & 1 &  99      &  1.545  &   54      & 15    \\
\  268   & 150.23084  & 2.5782007  & 19.64   & 0.08   & 0.86   & 1 &  99      &  1.402  &   51      & 15    \\
\  269   & 150.60793  & 2.5950421  & 23.25   & 0.12   & 0.00   & 2 &  95      &  0.627      &   42      &  0       \\
\  270   & 150.16382  & 2.5977140  & 22.23   & 0.24   & 0.96   & 1 &  99      & 1.589   &   51      & 15    \\
\  271   & 149.63532  & 2.5988504  & 22.02   & 0.06   & 0.83   & 1 &  97      & 2.536   &   53      & 15    \\
\  272   & 149.65574  & 2.6008206  & 21.63   & 0.12   & 0.03   & 1 &  99      & 0.735   &   54      & 15    \\
\  273   & 150.31792  & 2.6020259  & 21.30   & 0.20   & 0.56   & 1 &  99      & 0.958   &   54      & 15    \\
\  274   & 150.47164  & 2.6032998  & 22.01   & 0.11   & 0.70   & 2 &  98      & -       &   52      &  0       \\
\  275   & 150.54274  & 2.6059349  & 22.29   & 0.26   & 0.77   & 1 &  99      & 0.98    &   54      &  6      \\
\  276   & 149.85475  & 2.6069132  & 22.96  & 0.51    & 0.89   & 1 &  99      & 2.111$\diamond$  &   46      &  6      \\
\  277   & 149.73649  & 2.6107833  & 22.12  & 0.13    & 0.27   & 2 &  99      & 1.324   &   53      &  7     \\
\  278   & 150.59743  & 2.6179205  & 21.90  & 0.28    & 0.87   & 1 &  99      & 1.447   &   54      &  7     \\
\  279   & 150.59193  & 2.6193782  & 23.09  & 0.12    & 0.94   & 1 &  98      & 2.994$\triangle$       &   53      &  4      \\
\  280   & 150.12688  & 2.6265730   & 22.56  & 0.40   & 0.77   & 1 &  99      & 1.837   &   53      &  7     \\
\  281   & 150.04250  & 2.6291486  & 21.07  & 0.10    & 0.90   & 2 &  98      & 1.569   &   54      &  3      \\
\  282   & 150.04189  & 2.6294252  & 22.51   & 0.13   & 0.37   & 2 &  99      & -       &   54      &  2     \\
\  283   & 150.10281  & 2.6302075  & 22.54  & 0.08    & 0.04   & 2 &  95      & 1.182   &   53      &  3      \\
\  284   & 150.09495  & 2.6338103  & 23.30  & 0.17    & 0.94   & 1 &  99      & 1.183   &   51      &  2     \\
\  285   & 150.08056  & 2.6345305  & 22.42  & 0.08    & 0.98   & 1 &  95      & 0.71    &   53      & 15    \\
\  286   & 150.55491  & 2.6410387  & 21.97  & 0.08    & 0.38   & 1 &  98      & 1.142   &   54      & 15    \\
\  287   & 150.03431  & 2.6411103  & 22.05  & 0.07    & 0.04   & 1 &  97      & 0.518   &   54      & 15    \\
\  288   & 149.80802  & 2.6456649  & 21.63  & 0.23    & 0.90   & 1 &  99      & 2.082   &   54      &  7     \\
\  289   & 150.45662  & 2.6481900  & 20.67  & 0.05    & 0.87   & 1 &  97      & 2.05    &   54      & 15    \\
\  290   & 150.49762  & 2.6599360  & 20.31  & 0.17    & 0.88   & 1 &  99      & 0.854   &   54      & 15    \\
\  291   & 149.88379  & 2.6664010  & 22.54   & 0.08   & 0.66   & 1 &  95      & 1.346   &   54      &  6      \\ 
\  292   & 150.29776  & 2.6734549  & 22.16   & 0.20   & 0.91   & 1 &  99      & 1.593   &   54      &  6      \\
\  293   & 150.59885  & 2.6743665  & 20.81   & 0.04   & 0.02   & 2 &  95      & 0.066$\triangle$       &   54      &  0       \\
\  294   & 150.50550  & 2.6749128  & 21.79   & 0.27   & 0.97   & 1 &  99      & 1.289   &   54      &  6     \\
\  295   & 150.35149  & 2.6781381  & 20.96   & 0.31   & 0.85   & 1 &  99      & 2.754   &   54      & 15    \\
\  296   & 150.03067  & 2.6787261  & 22.30   & 0.37   & 0.92   & 1 &  99      & 1.951   &   54      &  7     \\
\  297   & 149.92542  & 2.6842236  & 20.57   & 0.21   & 0.90   & 1 &  99      & 1.779   &   53      &  3      \\
\  298   & 150.04527  & 2.6884792  & 22.08   & 0.19   & 0.98   & 1 &  99      & 1.436   &   54      &  7     \\
\  299   & 149.87593  & 2.6902237  & 21.78   & 0.06   & 0.87   & 1 &  97      & 2.169   &   54      & 15    \\
\hline
\end{longtable}
\end{landscape}


\end{document}